\documentclass[aps,prd,preprint,groupedaddress]{revtex4-1}
\usepackage[utf8]{inputenc}
\usepackage[english]{babel}
\usepackage[letterpaper,margin=1in]{geometry}

\usepackage[dvipsnames]{xcolor}
\usepackage{amsmath,amssymb,amsfonts,enumerate,float,graphicx,hyperref,mathrsfs,mathtools,fancyvrb,slashed,pdfpages,empheq,tabularx,caption,multirow,subcaption}

\usepackage[inline,shortlabels]{enumitem}

\allowdisplaybreaks

\usepackage{colortbl}
\usepackage{tikz,pgfplots}

\usetikzlibrary{calc,patterns,decorations.pathmorphing,decorations.markings,arrows}

\graphicspath{{./figures/}}

\usepackage[most]{tcolorbox}

\tcbset{
	frame code={}
	center title,
	left=0pt,
	right=0pt,
	top=0pt,
	bottom=0pt,
	colback=yellow!25,
	colframe=white,
	width=\dimexpr\textwidth\relax,
	enlarge left by=0mm,
	boxsep=5pt,
	arc=0pt,outer arc=0pt,
}

\newcommand\myshade{85}
\hypersetup{
	linkcolor  = violet!\myshade!black,
	citecolor  = YellowOrange!\myshade!black,
	urlcolor   = Aquamarine!\myshade!black,
	colorlinks = true,
}

\newcommand{\del}{\partial}

\makeatletter
\newcommand\incircbin{\mathpalette\@incircbin}
\newcommand\@incircbin[2]{\mathbin{\ooalign{\hidewidth$#1#2$\hidewidth\crcr$#1\bigcirc$}}}

\makeatother

\newcommand{\prg}{\\ $  $ \\  }

\newcommand{\ar}{\nonumber \\ & + & }
\newcommand{\ek}{\nonumber \\ & - & }
\newcommand{\carp}{\nonumber \\ & \times & }

\def\ba{\begin{eqnarray}}
\def\ea{\end{eqnarray}}
\def\bt{\begin{eqnarray*}}
	\def\et{\end{eqnarray*}}

\makeatletter
\newcommand{\pushright}[1]{\ifmeasuring@#1\else\omit\hfill$\displaystyle#1$\fi\ignorespaces}
\newcommand{\pushleft}[1]{\ifmeasuring@#1\else\omit$\displaystyle#1$\hfill\fi\ignorespaces}
\makeatother

\def\mqi{m_{q_{1}}}
\def\mqii{m_{q_{2}}}
\def\mV{m_{V}}
\def\fV{f_{V}}
\def\fVT{f_{V}^{T}}
\def\baru{\bar{u}}
\def\alphai{\alpha_{1}}
\def\alphaiii{\alpha_{3}}
\def\aipar{a_{1}^{\parallel}}
\def\aiper{a_{1}^{\perp}}
\def\aiipar{a_{2}^{\parallel}}
\def\aiiper{a_{2}^{\perp}}
\def\zetaiiipar{\zeta_{3}^{\parallel}}
\def\lambdaiiipartilde{\tilde{\lambda}_{3}^{\parallel}}
\def\omegaiiipar{\omega_{3}^{\parallel}}
\def\omegaiiipartilde{\tilde{\omega}_3^\parallel}
\def\lambdaiiipar{\lambda_{3}^{\parallel}}
\def\kappaiiipar{\kappa_{3}^{\parallel}}
\def\kappaiiiper{\kappa_{3}^{\perp}}
\def\omegaiiiper{\omega_{3}^{\perp}}
\def\lambdaiiiper{\lambda_{3}^{\perp}}
\def\zetaivpar{\zeta_{4}^{\parallel}}
\def\omegaivpartilde{\tilde{\omega}_{4}^{\parallel}}
\def\zetaivper{\zeta_{4}^{\perp}}
\def\zetaivpertilde{\tilde{\zeta}_{4}^{\perp}}
\def\kappaivpar{\kappa_{4}^{\parallel}}
\def\kappaivper{\kappa_{4}^{\perp}}
\def\thetaipar{\theta_{1}^{\parallel}}
\def\thetaiipar{\theta_{2}^{\parallel}}
\def\psiiipar{\psi_{2}^{\parallel}}
\def\thetaiper{\theta_{1}^{\perp}}
\def\thetaipertilde{\tilde{\theta}_{1}^{\perp}}
\def\thetaiiper{\theta_{2}^{\perp}}
\def\thetaiipertilde{\tilde{\theta}_{2}^{\perp}}
\def\phiiipertilde{\tilde{\phi}_{2}^{\perp}}
\def\phiiiper{\phi_{2}^{\perp}}
\def\avqi{\langle\langle Q^{(1)}\rangle\rangle}
\def\avqiii{\langle\langle Q^{(3)} \rangle\rangle}
\def\avqv{\langle\langle Q^{(5)} \rangle\rangle}
\def\psioper{\psi_{0}^{\perp}}
\def\psiopertilde{\tilde{\psi}_{0}^{\perp}}
\def\thetaoper{\theta_{0}^{\perp}}
\def\thetaopertilde{\tilde{\theta}_{0}^{\perp}}
\def\phioper{\phi_{0}^{\perp}}
\def\phiopertilde{\tilde{\phi}_{0}^{\perp}}
\def\phiiper{\phi_{1}^{\perp}}
\def\phiipertilde{\tilde{\phi}_{1}^{\perp}}
\def\psiiper{\psi_{1}^{\perp}}
\def\psiipertilde{\tilde{\psi}_{1}^{\perp}}
\def\psiiiper{\psi_{2}^{\perp}}
\def\psiiipertilde{\tilde{\psi}_{2}^{\perp}}
\def\BorelM2{M^2}
\def\NN{N^2}
\def\mQ{m_Q}
\def\mQi{m_{Q'}}

\def\phiiiparu{\phi _{2} ^{\parallel} (u)}
\def\phiiiperu{\phi _{2} ^{\perp} (u)}
\def\phiiiiparu{\phi _{3} ^{\parallel} (u)}
\def\psiiiiparu{\psi _{3} ^{\parallel} (u)}
\def\psiiiiperu{\psi _{3} ^{\perp} (u)}
\def\phiiiiperu{\phi _{3} ^{\perp} (u)}
\def\psiivparu{\psi _{4} ^{\parallel} (u)}
\def\phiivparu{\phi _{4} ^{\parallel} (u)}
\def\psiivperu{\psi _{4} ^{\perp} (u)}
\def\phiivperu{\phi _{4} ^{\perp} (u)}
\def\SS{\mathcal{S} (\alpha _{1}, \alpha _{3})}
\def\SSTilde{\tilde{\mathcal{S}} (\alpha _{1}, \alpha _{3})}
\def\VV{\mathcal{V} (\alpha _{1}, \alpha _{3})}
\def\AA{\mathcal{A} (\alpha _{1}, \alpha _{3})}
\def\TT{\mathcal{T} (\alpha _{1}, \alpha _{3})}
\def\TTi{\mathcal{T} _{1}^{(4)} (\alpha _{1}, \alpha _{3})}
\def\TTii{\mathcal{T} _{2}^{(4)} (\alpha _{1}, \alpha _{3})}
\def\TTiii{\mathcal{T} _{3}^{(4)} (\alpha _{1}, \alpha _{3})}
\def\TTiv{\mathcal{T} _{4}^{(4)} (\alpha _{1}, \alpha _{3})}

\def\Hat{\mathcal H}
\def\Der{\mathcal D}
\def\I{\mathcal I}
\def\Ixx{\mathcal I _2}

\def\IG{\mathcal G}
\def\IGxx{\mathcal G _2}

\makeatletter
\AtBeginDocument{\let\LS@rot\@undefined}
\makeatother

\begin{document}
\title{Strong coupling constants of doubly heavy baryons with vector mesons in QCD}

\author{T. M. Aliev}
\email[]{taliev@metu.edu.tr}
\affiliation{Physics Department, Middle East Technical University, Ankara 06800, Turkey}

\author{K. \c Sim\c sek}
\email[]{ksimsek@u.northwestern.edu}
\affiliation{Department of Physics \& Astronomy, Northwestern University, Evanston, IL 60208, USA}

\date{September 7, 2020}

\begin{abstract}
	Using the most general form of the interpolating current for baryons, the strong electric and magnetic coupling constants of light vector mesons $ \rho $ and $ K^* $ with doubly heavy baryons are computed within the light-cone sum rules. We consider 2- and 3-particle distribution amplitudes of the aforementioned vector mesons. The obtained results can be useful in the analysis of experimental data on the properties of doubly heavy baryons conducted at LHC.
\end{abstract}
%\begin{center}
%\LARGE\textbf{Strong coupling constants of doubly heavy baryons with vector mesons in QCD}
%\prg \large
%T. M. Aliev\footnote{email: \href{mailto:taliev@metu.edu.tr}{\textsf{taliev@metu.edu.tr}}}\\\textit{Department of Physics, Middle East Technical University, Ankara 06800, Turkey}
%\prg
%K. \c Sim\c sek\footnote{email: \href{mailto:ksimsek@u.northwestern.edu}{\textsf{ksimsek@u.northwestern.edu}}}\\\textit{Department of Physics \& Astronomy, Northwestern University, Evanston 60208, IL, USA}
%\prg
%September 6, 2020
%\end{center}
%$ $\\ 
\maketitle 
\section{Introduction}
The quark model has been very predictive in studying the properties of hadrons \cite{ref:onceki-1}. Many baryon states predicted by the quark model have already been observed experimentally. 
\prg 
The spectroscopy of doubly heavy baryons has been extensively investigated in many theoretical works. It has been studied in the framework of different approaches such as the Hamilton method \cite{ref:onceki-9}, the hypercentral method \cite{ref:onceki-10}, the lattice QCD \cite{ref:onceki-11, ref:onceki-12}, the QCD sum rules \cite{ref:onceki-13, ref:onceki-14, ref:onceki-15, ref:onceki-16, ref:onceki-17, ref:onceki-18}, the Bethe-Salpeter equation \cite{ref:onceki-19}, and in an extended chromomagnetic model \cite{ref:onceki-20}.
\prg 
At the present time, only the $ \Xi _{cc}$ state is observed in experiments. This state was first observed by SELEX Collaboration in $ \Xi _{cc} \to \Lambda _{c} K^- \pi ^+ $ and $ p D^+ K^- $ reactions with the mass $ 3518.7 \pm 1.7 $ MeV \cite{ref:aliev-1, ref:aliev-2}. In 2017, the doubly heavy $ \Xi _{cc} ^{++} $ was discovered by LHCb Collaboration with the measured mass of $ 3621.24 \pm 0.65 \pm 0.31 $ MeV \cite{ref:aliev-3}. The LHCb Collaboration also measured the lifetime of this state: $ \tau = 0.256 {\tiny\begin{matrix}
+0.024\\ -0.022
\end{matrix}} \pm 0.014 $ ps \cite{ref:aliev-4}. 
\prg 
This discovery is stimulated by many theoretical works for a deeper understanding of the properties of these baryons via studying their electromagnetic, weak, and strong decays. The weak decays of the doubly heavy baryons have been studied within various approaches such as the light-front approach \cite{ref:aliev-5, ref:aliev-9}, the QCD sum rules approach \cite{ref:aliev-6}, the quark model \cite{ref:aliev-7}, and the covariant light-front quark model \cite{ref:aliev-8}.
\prg
The radiative transitions of doubly heavy baryons in the framework of different approaches such as relativized quark model \cite{ref:aliev-10}, in the chiral perturbation theory \cite{ref:aliev-11}, in the light-cone QCD sum rules \cite{ref:aliev-12} are comprehensively studied. The strong coupling constant of light pseudoscalar mesons with doubly heavy baryons within the light-cone QCD sum rules are studied in \cite{ref:aliev-13} and \cite{ref:aliev-14}.  
\prg 
In this work, we study the strong coupling constants of doubly heavy baryons with vector mesons $ \rho $ and $ K^* $ within the framework of the light-cone sum rules (LCSR). These coupling constants can play an important role in the description of doubly heavy baryons in terms of one boson exchange potential models. %There have been numerous studies in the literature on this subject. To illustrate, the $ NN\rho $ strong coupling constant is studied by using the LCSR method in \cite{ref:eski-4}. Also, the strong coupling constants of $ NN\rho $, $ \Sigma \Sigma \rho $, and $ \Xi \Xi \rho $ are studied in the LCSR framework in \cite{ref:eski-5}. Furthermore, the coupling constant of the vector mesons $ \rho $ and $ \omega $ with the baryons is studied in the framework of the external field sum rules method in \cite{ref:eski-6}. We kindly refer the reader to \cite{ref:eski-8, ref:eski-9, ref:onceki-28} for a detailed discussion on the LCSR method. 
\prg 
This paper is organized as follows. In Section \ref{sec:2}, we derive the LCSR for the electric- and magnetic-type strong coupling constants of doubly heavy baryons with vector mesons $ \rho $ and $ K^* $, as well as present the details of the calculations. Section \ref{sec:3} is devoted to the numerical analysis of the sum rules for the said coupling constants. Section \ref{sec:4} contains our conclusion.  
\section{Light-cone sum rules for vector meson-baryon coupling constants}\label{sec:2}
For determining the strong coupling constants of doubly heavy baryons with light vector mesons, we introduce the following correlation function:
\ba
  \Pi = i \int d^4x \ e^{ipx} \langle V(q) \vert T\{ \eta(x) \bar\eta(0) \} \vert 0 \rangle \label{eq:1}
\ea
where $ V(q) $ is a vector meson with momentum $ q $ and $ \eta $ denotes the interpolating current of the corresponding doubly heavy baryon. By the virtue of the $ SU(3) $ classification, there exist two types of interpolating currents, which are symmetric or antisymmetric under the exchange of two heavy quarks. Only when the two heavy quarks are different, we have the antisymmetric current. The most general forms of the interpolating currents, both symmetric and antisymmetric, for doubly heavy baryons with $ J = 1/2 $ can be written as
\ba
  \eta ^{(S)} = \frac{1}{\sqrt 2} \epsilon ^{abc} \sum _{i=1}^2 [ ({Q^a}^{\rm T} A _1^i q^b) A_2^i {Q'}^c + (Q \leftrightarrow Q') ]
\ea
and
\ba
  \eta ^{(A)} = \frac{1}{\sqrt 6} \epsilon ^{abc} \sum _{i=1}^2 [ 2({Q^a}^{\rm T} A_1^i {Q'}^b) A_2^i q^c + ({Q^a}^{\rm T} A_1^i q^b) A_2^i {Q'}^c - ({{Q'}^a}^{\rm T} A_1^i q^b) A_2^i Q^c ]
\ea
where $ a $, $ b $, and $ c $ are color indices, $ \rm{T} $ is the transposition, and 
\ba
  A_1^1 = C,\quad A_1^2 = C\gamma_5,\quad A_2^1 = \gamma_5,\quad A_2^2 = \beta I
\ea
where $ \beta $ is an arbitrary parameter and $ C $ is the charge conjugation operator.
\prg 
The main philosophy of the light-cone sum rules (LCSR) is the computation of the correlation function in two different domains. It can be calculated in terms of the hadrons, as well as in the deep Euclidean region $ p^2 \to -\infty $ and $ (p+q)^2 \to -\infty $ by using the operator product expansion (OPE) over twist. Afterwards, the corresponding double Borel transformation with respect to the variables $ -p^2 $ and $ -(p+q)^2 $ is performed to suppress the contributions from higher states and the continuum as well as to enhance the contributions by the ground state. Finally, matching the results, the desired sum rules is obtained. 
\prg 
We start the construction of the sum rules by considering the phenomenological part of the correlation function. To this end, we insert a complete set of intermediate states with the same quantum numbers as the interpolating currents. After isolating the ground-state baryons, we obtain
\ba
  \Pi = \frac{\langle 0 \vert \eta \vert B_2 (p_2) \rangle \langle B_2 (p_2) V(q) \vert B_1 (p_1) \rangle \langle B_1 (p_1) \vert \bar\eta \vert 0 \rangle}{(p_2^2 - m_{B_2}^2) (p_1^2 - m_{B_1}^2)} + \mbox{higher states} \label{eq:5}
\ea
where $ m_{B_1} $ and $ m_{B_2} $ are the masses of the initial and final doubly heavy baryons, respectively. The matrix elements in Eq. \eqref{eq:5} are defined as follows:
\ba
  \langle 0 \vert \eta \vert B _i (p_i) \rangle &=& \lambda _{B _i} u(p_i) \label{eq:6} \\
  \langle B_2 (p_2) V(q) \vert B_1 (p_1) \rangle &=& \bar u (p_2) (f_1 \gamma_\mu - f_2 \frac{i}{m_{B_1} + m_{B_2}} \sigma_{\mu\nu}q^\nu) u(p_1) \varepsilon^\mu \label{eq:7}
\ea
where the $ \lambda _{B_i} $ are the residues, $ f_1 $ and $ f_2 $ are the relevant coupling constants of the doubly heavy baryons with the corresponding vector meson, $ \varepsilon _\mu$ and $ q_\mu $ are the 4-polarization and 4-momentum of the vector meson, and $ u $ is the Dirac spinor for the baryon which is normalized as $ \bar u u = 2m $. 
\prg
Using Eqs. \eqref{eq:6} and \eqref{eq:7} in \eqref{eq:5}, we obtain the following for the physical part of the correlation function:
\ba
  \Pi ^{\rm phys} &=& \frac{\lambda _{B_1} \lambda_{B_2}}{(p_1^2-m_{B_1}^2) (p_2^2-m_{B_2}^2)} \varepsilon^\mu (\slashed p_2 +m_{B_2}) (f_1 \gamma_\mu - f_2 \frac{i}{m_{B_1} + m_{B_2}} \sigma_{\mu\nu} q^\nu) (\slashed p_1 + m_{B_1}) \nonumber \\
  &=&  \frac{\lambda_{B_1} \lambda_{B_2}}{(p^2-m_{B_2}^2) ((p+q)^2 - m_{B_1}^2)} (
    \slashed p \slashed \varepsilon \slashed q (f_1 + f_2) 
    + 2(\varepsilon \cdot p) \slashed p f_1 + \mbox{other structures} )  \label{eq:8}
\ea
where we have set $ p_1 = p+q $ and $ p_2 = p $. 
\prg 
The correlation function given by Eq. \eqref{eq:8} contains many structures. Our numerical analysis shows that these structures give more reliable results and hence we choose them. % Performing a Borel transformation over variables $ p^2 $ and $ -(p+q)^2 $ and the continuum subtraction, the sum rules for the electric and magnetic type coupling constants are obtained as 
On the other hand, the correlation function is  calculated from the QCD side by using the OPE over twist. After applying the Wick theorem, from \eqref{eq:1}, we get the following results:
\ba 
	\Pi ^{(SS)} &=& \frac 12 \epsilon ^{abc} \epsilon^{a'b'c'} \int d^4x \ e^{ip\cdot x} \sum _{ij} (A_1^i) _{\alpha\beta} (A_2^i) _{\rho\gamma} (\tilde A_2^j) _{\gamma' \rho'} (\tilde A _1^j) _{\alpha'\beta'}
	\carp \langle V(q) \vert [S _{Q'\gamma\gamma'}^{cc'} S_{Q\alpha\beta'}^{aa'} + (Q\leftrightarrow Q') - S_{Q\alpha\gamma'}^{ac'} S_{Q'\gamma\beta'}^{ca'} - (Q \leftrightarrow Q')] q_\beta^b \bar q_{\alpha'}^{b'} \vert 0 \rangle \label{eq:11}
\ea 
\ba 
	\Pi ^{(AA)} &=& \frac 16 \epsilon ^{abc} \epsilon ^{a'b'c'} \int d^4x\ e^{ip\cdot x} \sum _{ij} (A_1^i) _{\alpha\beta} (A_2^i) _{\rho\gamma} (\tilde A_2^j) _{\gamma'\rho'} (\tilde A_1^j) _{\alpha'\beta'} 
	\carp \langle V(q) \vert 
	4 S _{Q'\beta\alpha'} ^{bb'} S _{Q\alpha\beta'}^{aa'} q _\gamma ^c \bar q _{\gamma'} ^{c'}
	-2 S_{Q'\beta\gamma'}^{bc'} S_{Q\alpha\beta'}^{aa'} q_\gamma^c \bar q_{\alpha'}^{b'} 
	- 2 S_{Q\alpha\gamma'}^{ac'} S_{Q'\beta\beta'}^{ba'} q_\gamma ^c \bar q _{\alpha'}^{b'}
	\ek 2 S_{Q'\gamma \alpha'}^{cb'} S_{Q\alpha\beta'}^{aa'} q _\beta ^b \bar q_{\gamma'} ^{c'} 
	+ S_{Q'\gamma\gamma'} ^{cc'} S_{Q\alpha\beta'} ^{aa'} q_\beta^b \bar q _{\alpha'} ^{b'} 
	+ q_\beta^b \bar q_{\alpha'} ^{b'} S _{Q\alpha\gamma'} ^{ac'} S_{Q'\gamma\beta'} ^{ca'} 
	- 2 q _\beta ^b \bar q _{\gamma'} ^{c'} S _{Q'\alpha\alpha'} ^{ab'} S _{Q\gamma\beta'} ^{ca'} 
	\ar q _\beta ^b \bar q _{\alpha'} ^{b'} S _{Q'\alpha\gamma' }^{ac'} S _{Q\gamma\beta'} ^{ca'} 
	+ q _\beta ^b \bar q _{\alpha'} ^{b'} S _{Q\gamma\gamma' }^{cc'} S _{Q'\alpha\beta' }^{aa'} 
	\vert 0 \rangle \label{eq:12}
\ea 
\ba 
	\Pi ^{(SA)} &=& \frac{1}{\sqrt{12}} \epsilon^{abc} \epsilon^{a'b'c'} \int d^4x\ e^{ip\cdot x} \sum _{ij} (A_1^i) _{\alpha\beta} (A_2^i) _{\rho\gamma} (\tilde A_2^j) _{\gamma'\rho'} (\tilde A_1^j) _{\alpha'\beta'} 
	\carp \langle V(q) \vert 
		-2 q_\gamma ^c \bar q _{\alpha'} ^{b'} S_{Q'\beta\gamma'}^{bc'} S_{Q\alpha\beta'} ^{aa'} 
		+ 2 q _\gamma ^c \bar q _{\alpha'} ^{b'} S _{Q\alpha\gamma'} ^{ac'} S _{Q'\beta\beta'} ^{ba'}
		+ q _\beta ^b \bar q_{\alpha'} ^{b'} S_{Q'\gamma\gamma'} ^{cc'} S _{Q\alpha\beta'} ^{aa'}
		\ek q _\beta ^b \bar q _{\alpha'} ^{b'} S _{Q\alpha\gamma'} ^{ac'} S _{Q'\gamma\beta'} ^{ca'} 
		+ q _\beta ^b \bar q_{\alpha'} ^{b'} S _{Q'\alpha\gamma'} ^{ac'} S _{Q\gamma\beta'} ^{ca'}
		- q _\beta ^b \bar q_{\alpha'}^{b'} S _{Q'\alpha\beta'} ^{aa'} S _{Q\gamma\gamma'} ^{cc'}
	\vert 0 \rangle \label{eq:13}
\ea 
where $ S _Q$ is the heavy quark propagator. In these expressions, the superscripts $ (SS) $, $ (AA) $, and $ (SA) $ denote the symmetry property of the currents $ \eta $ and $ \bar \eta $, and we have defined $ \tilde A _i = \gamma ^0 A _i ^\dagger \gamma ^0 $. The heavy quark propagator in the presence of an external background field in the coordinate space is given by
\ba 
	S _{Q\alpha\beta} ^{aa'} &=& \frac{m_Q^2}{4\pi} (\frac{iK _2(m _Q \sqrt{-x^2})\slashed x}{(\sqrt{-x^2})^2} + \frac{K _1(m _Q \sqrt{-x^2})}{\sqrt{-x^2}}) _{\alpha\beta} \delta ^{aa'} 
	\ek \frac{g _s}{16\pi^2} m _Q \int _0 ^1 du\ (\frac{iK _1(m _Q \sqrt{-x^2})}{\sqrt{-x^2}} (\bar u \slashed x \sigma _{\lambda \tau} + u \sigma _{\lambda \tau} \slashed x) + K _0 (m _Q\sqrt{-x^2})\sigma _{\lambda \tau}) _{\alpha\beta} 
	\carp G ^{(n)\lambda \tau} (\frac{\lambda ^{(n)}}{2})^{aa'} \label{eq:14}
\ea 
where $ G _{\lambda \tau} ^{(n)} $ is the gluon field strength tensor, the $ \lambda ^{(n)} $ are the Gell-Mann matrices, and the $ K _i(m _Q \sqrt{-x^2}) $ are the modified Bessel functions of the second kind. Using the Fiertz identities, namely
\ba 
	q _\alpha ^b \bar q _{\beta} ^{b'} \to -\frac{1}{12} (\Gamma _i) _{\alpha\beta} \delta ^{bb'} [\bar q \Gamma _i q] \label{eq:15}
\ea 
and
\ba 
	q _\alpha ^b \bar q _\beta ^{b'} G _{\lambda\tau} ^{(n)} \to -\frac{1}{16} (\frac{\lambda ^{(n)}}{2}) ^{bb'} (\Gamma _i) _{\alpha\beta} [\bar q \Gamma _i G _{\lambda \tau} ^{(n)} q] \label{eq:16}
\ea 
one can see that the following matrix elements appear in the calculation:
\ba 
	\langle V(q,\varepsilon) \vert \bar q \Gamma _i q \vert 0 \rangle \quad \mbox{and} \quad \langle V(q,\varepsilon) \vert \bar q \Gamma _i G _{\mu\nu} ^{(n)} q \vert 0 \rangle 
\ea 
$ \{ \Gamma _i \} $ is the full set of Dirac matrices, i.e.  
\ba 
	\Gamma _1 = I, \  \Gamma _2 = \gamma _5, \ \Gamma _3 = \gamma _\alpha, \ \Gamma _4 = i \gamma _\alpha \gamma _5,\ \Gamma _5 = \frac{1}{\sqrt 2} \sigma _{\alpha\beta}
\ea 
Now if we insert Eqs. \eqref{eq:14} -- \eqref{eq:16} into Eqs. \eqref{eq:11} -- \eqref{eq:13}, do calculations for the QCD part of the correlation function, and perform a double Borel transformation over variables $ -p^2 $ and $ -(p+q)^2 $, we obtain the following results for the considered structures:
\ba 
	\Pi ^{(SS){\rm theo}}_{f_1+f_2} &=& 1/(96 \BorelM2 (\pi)^4) i \mQ \mQi \NN (-(-1+(\beta)^2) \fVT (3 (\mQ+\mQi) \mV \IG[\SS,0,2]
	\ar 9 (\mQ+\mQi) \mV \IG[\SS,1,1]-3 \mQ \mV (i \IG[\SSTilde,0,2]
	\ar \fVT \mV (\IG[\TTiii,1,1]-\IG[\TTiv,1,1]-2 \IG[\TTiii u,1,1]
	\ar 2 \IG[\TTiv u,1,1]+(\mV)^2 (\IGxx[\Hat _G[2,\TTi],1,1]
	\ek 2 \IGxx[u \Hat _G[2,\TTi],1,1]-\IGxx[\Hat _G[2,\TTii],1,1]
	\ar 2 \IGxx[u \Hat _G[2,\TTii],1,1])))-\mQi (224 \mQ \I[\phiiiperu,1,2]
	\ar \mV (3 i \IG[\SSTilde,0,2]+\mV (3 \fVT (\IG[\TTiii,1,1]-\IG[\TTiv,1,1]
	\ek 2 \IG[\TTiii u,1,1]+2 \IG[\TTiv u,1,1]
	\ar (\mV)^2 (\IGxx[\Hat _G[2,\TTi],1,1]-2 \IGxx[u \Hat _G[2,\TTi],1,1]
	\ek \IGxx[\Hat _G[2,\TTii],1,1]+2 \IGxx[u \Hat _G[2,\TTii],1,1]))
	\ar 14 \mQ \Ixx[\phiivperu,1,2]))))+\fV i \mV (3 (1+\beta (6+\beta)) (\mQ+\mQi) 
	\carp \IG[\Der _G[1,\AA],1,2]-6 (1+\beta (6+\beta)) (\mQ+\mQi) 
	\carp \IG[u \Der _G[1,\AA],1,2]-3 (3+\beta (2+3 \beta)) \mQ \IG[\Der _G[1,\VV],1,2]
	\ar \mQi (-3 (3+\beta (2+3 \beta)) \IG[\Der _G[1,\VV],1,2]
	\ek 28 i \mQ ((-1+\beta)^2 \I[\psiiiiperu,1,1]-(1+\beta)^2 \Ixx[\psiiiiperu,2,2])))) \label{eq:17}
\ea 
\ba 
\Pi ^{(SS){\rm theo}}_{f_1} &=& -1/(96 (\BorelM2)^2 (\pi)^4) i \mQ \mQi \mV \NN (3 i (-1+(\beta)^2) (\fVT)^2 (\mQ+\mQi) (\mV)^3 
\carp (6 \IG[\Hat _G[1,\TT],0,2]+2 (\IG[\Hat _G[1,\TT],1,1]
\ek \IG[\Hat _G[1,\TTi],0,2]+\IG[\Hat _G[1,\TTi],1,1]
\ar \IG[\Hat _G[1,\TTii],0,2]+4 \IG[\Hat _G[1,\TTiii],0,2]
\ar 2 \IG[\Hat _G[1,\TTiv],0,2]+\IG[\Hat _G[1,\TTiv],1,1])
\ar 3 (\mV)^2 \IGxx[\Hat _G[3,\TT],0,2])-2 \fV (3 (3+\beta (2+3 \beta)) (\mQ+\mQi) (\mV)^2 
\carp \IG[\AA,1,2]-3 (3+\beta (2+3 \beta)) (\mQ+\mQi) (\mV)^2 \IG[\VV,1,2]
\ar 6 (3+\beta (2+3 \beta)) (\mQ+\mQi) (\mV)^2 \IG[\VV u,1,2]
\ek 14 \mQ \mQi (16 (3+\beta (2+3 \beta)) \I[\phiiiparu,2,2]+(\mV)^2 (-8 (-1+\beta)^2 
\carp \I[\Hat[2,\phiiiiperu],1,1]+4 (-1+\beta)^2 \I[\Hat[2,\phiiiparu],1,1]+4 (-1+\beta)^2 
\carp \I[\Hat[2,\psiivparu],1,1]+(3+\beta (2+3 \beta)) \Ixx[\phiivparu,2,2]-4 (5+\beta (2+5 \beta)) 
\carp (-2 \Ixx[\Hat[2,\phiiiiperu],2,2]+\Ixx[\Hat[2,\phiiiparu],2,2]+\Ixx[\Hat[2,\psiivparu],2,2]))))) \label{eq:18}
\ea
\ba 
\Pi ^{(AA){\rm theo}}_{f_1+f_2} &=& (1/(576 \BorelM2 (\pi)^4))\mQ \mQi ((-1+\beta) \fVT i (3 (5+\beta) (\mQ+\mQi) \mV 
\carp \IG[\SS,0,2]-3 (5+\beta) (\mQ+\mQi) \mV \IG[\SS,1,1]+(1+5 \beta) 
\carp (-3 i (\mQ+\mQi) \mV \IG[\SSTilde,0,2]-3 \fVT (\mQ+\mQi) (\mV)^2 
\carp (\IG[\TTiii,1,1]-\IG[\TTiv,1,1]-2 \IG[\TTiii u,1,1]
\ar 2 \IG[\TTiv u,1,1]+(\mV)^2 (\IGxx[\Hat _G[2,\TTi],1,1]
\ek 2 \IGxx[u \Hat _G[2,\TTi],1,1]-\IGxx[\Hat _G[2,\TTii],1,1]
\ar 2 \IGxx[u \Hat _G[2,\TTii],1,1]))-14 \mQ \mQi (16 \I[\phiiiperu,1,2]
\ar (\mV)^2 \Ixx[\phiivperu,1,2])))+2 \fV \mV (6 ((1+\beta+(\beta)^2) \mQ-\beta \mQi) 
\carp \IG[\Der _G[1,\AA],1,2]+3 (\mQ+\beta (6+\beta) \mQ-(-1+\beta)^2 \mQi) 
\carp \IG[u \Der _G[1,\AA],1,2]-3 (1+\beta (4+\beta)) \mQ \IG[\Der _G[1,\VV],1,2]
\ar \mQi (3 (1+(\beta)^2) \IG[\Der _G[1,\VV],1,2]-14 i \mQ ((-1+\beta) (11+13 \beta) 
\carp \I[\psiiiiperu,1,1]-(13+\beta (10+13 \beta)) \Ixx[\psiiiiperu,2,2])))) \label{eq:19}
\ea 
\ba 
\Pi ^{(AA){\rm theo}}_{f_1} &=& (1/(576 (\BorelM2)^2 (\pi)^4))\mQ \mQi \mV (3 (-1+\beta) (\fVT)^2 (\mQ+\mQi) (\mV)^3 (2 (5+\beta) 
\carp \IG[\Hat _G[1,\TT],0,2]-2 (5+\beta) (\IG[\Hat _G[1,\TT],1,1]
\ek \IG[\Hat _G[1,\TTi],0,2]+\IG[\Hat _G[1,\TTi],1,1])
\ek 2 (1+5 \beta) \IG[\Hat _G[1,\TTii],0,2]+12 (1+\beta) \IG[\Hat _G[1,\TTiv],0,2]
\ek 2 (5+\beta) \IG[\Hat _G[1,\TTiv],1,1]+(1+5 \beta) (\mV)^2 
\carp \IGxx[\Hat _G[3,\TT],0,2])+2 i \fV (3 (-1+\beta)^2 (\mQ+\mQi) (\mV)^2 
\carp \IG[\AA,1,2]-3 (-1+\beta)^2 (\mQ+\mQi) (\mV)^2 \IG[\VV,1,2]
\ar 6 (-1+\beta)^2 (\mQ+\mQi) (\mV)^2 \IG[\VV u,1,2]-14 \mQ \mQi (16 (-1+\beta)^2 
\carp \I[\phiiiparu,2,2]+(\mV)^2 (-8 (-1+\beta) (13+11 \beta) \I[\Hat[2,\phiiiiperu],1,1]+4 (-1+\beta) 
\carp (13+11 \beta) \I[\Hat[2,\phiiiparu],1,1]-52 \I[\Hat[2,\psiivparu],1,1]+4 \beta (2+11 \beta) 
\carp \I[\Hat[2,\psiivparu],1,1]+(-1+\beta)^2 \Ixx[\phiivparu,2,2]-60 (-2 \Ixx[\Hat[2,\phiiiiperu],2,2]
\ar \Ixx[\Hat[2,\phiiiparu],2,2]+\Ixx[\Hat[2,\psiivparu],2,2])-12 \beta (2+5 \beta) 
\carp (-2 \Ixx[\Hat[2,\phiiiiperu],2,2]+\Ixx[\Hat[2,\phiiiparu],2,2]+\Ixx[\Hat[2,\psiivparu],2,2]))))) \label{eq:20}
\ea
\ba 
\Pi ^{(SA){\rm theo}}_{f_1+f_2} &=& (1/(32 \sqrt 3 \BorelM2 (\pi)^4))\mQ (\mQ-\mQi) \mQi \mV N 
\carp (2 \fV (-2 \beta \IG[\Der _G[1,\AA],1,2]+2 (1+\beta (4+\beta)) \IG[u \Der _G[1,\AA],1,2]
\ar (1+(\beta)^2) \IG[\Der _G[1,\VV],1,2]+3 (1+\beta)^2 \IG[u \Der _G[1,\VV],1,2])
\ar (-1+\beta) \fVT (-2 i \beta \IG[\SS,0,2]-2 i (1+2 \beta) \IG[\SS,1,1]
\ek 2 \IG[\SSTilde,0,2]-(-1+\beta) (\IG[\SSTilde,1,1]-2 \IG[\SSTilde u,1,1])
\ar \fVT i \mV ((1+3 \beta) \IG[\TTi,0,2]+(1+3 \beta) \IG[\TTi,1,1]
\ek \IG[\TTii,0,2]-\IG[\TTii,1,1]+3 \IG[\TTiii,0,2]
\ek 2 \IG[\TTiii,1,1]-3 \IG[\TTiv,0,2]-\beta (3 \IG[\TTii,0,2]
\ar 3 \IG[\TTii,1,1]-\IG[\TTiii,0,2]+\IG[\TTiv,0,2])
\ar 2 \IG[\TTiv,1,1]+4 \IG[\TTiii u,1,1]-4 \IG[\TTiv u,1,1])
\ar \fVT i (\mV)^3 ((3+\beta) \IGxx[\Hat _G[2,\TTi],0,2]-2 \IGxx[\Hat _G[2,\TTi],1,1]
\ar 4 \IGxx[u \Hat _G[2,\TTi],1,1]-(3+\beta) \IGxx[\Hat _G[2,\TTii],0,2]
\ar 2 (\IGxx[\Hat _G[2,\TTii],1,1]-2 \IGxx[u \Hat _G[2,\TTii],1,1])))) \label{eq:21}
\ea 
\ba 
\Pi ^{(SA){\rm theo}}_{f_1} &=& -(1/(96 \sqrt 3 (\BorelM2)^2 (\pi)^4))\mQ \mQi \mV N (3 (-1+\beta) (\fVT)^2 (\mQ-\mQi) (\mV)^3 
\carp ((4+8 \beta) \IG[\Hat_G[1,\TT],0,2]+2 (2 \beta \IG[\Hat _G[1,\TT],1,1]
\ar 2 \beta \IG[\Hat _G[1,\TTi],0,2]+2 \beta \IG[\Hat _G[1,\TTi],1,1]
\ek 2 \IG[\Hat _G[1,\TTii],0,2]+3 \IG[\Hat _G[1,\TTii],1,1]
\ar \beta \IG[\Hat _G[1,\TTii],1,1]-6 \IG[u \Hat _G[1,\TTii],1,1]
\ek 2 \beta \IG[u \Hat _G[1,\TTii],1,1]+2 \IG[\Hat _G[1,\TTiii],0,2]
\ar 2 \beta \IG[\Hat _G[1,\TTiii],0,2]+4 \IG[\Hat _G[1,\TTiv],0,2]
\ar 4 \beta \IG[\Hat _G[1,\TTiv],0,2]-3 \IG[\Hat _G[1,\TTiv],1,1]
\ar \beta \IG[\Hat _G[1,\TTiv],1,1]+2 (3+\beta) \IG[u \Hat _G[1,\TTiv],1,1])
\ar (\mV)^2 (2 (2+\beta) \IGxx[\Hat _G[3,\TT],0,2]
\ek (3+\beta) (\IGxx[\Hat _G[3,\TT],1,1]-2 \IGxx[u \Hat _G[3,\TT],1,1])))
\ar 2 i \fV (3 (-1+\beta)^2 (\mQ-\mQi) (\mV)^2 \IG[\AA,1,2]+18 (1+\beta)^2 
\carp (\mQ-\mQi) (\mV)^2 \IG[\AA u,1,2]-3 \mQ (\mV)^2 ((-1+\beta)^2 
\carp \IG[\VV,1,2]-8 (1+\beta+(\beta)^2) \IG[\VV u,1,2])+\mQi (3 (\mV)^2 
\carp ((-1+\beta)^2 \IG[\VV,1,2]-8 (1+\beta+(\beta)^2) \IG[\VV u,1,2])
\ar 14 (1+(\beta)^2) \mQ (16 \I[\phiiiparu,2,2]+(\mV)^2 (\Ixx[\phiivparu,2,2]
\ar 24 (-2 \Ixx[\Hat[2,\phiiiiperu],2,2]+\Ixx[\Hat[2,\phiiiparu],2,2]
\ar \Ixx[\Hat[2,\psiivparu],2,2])))))) \label{eq:22}
\ea
where $ N $ is the normalization factor which is equal to $ \frac{1}{\sqrt 2} $ (1) for different (identical) heavy quark flavors. In Eqs. \eqref{eq:17} -- \eqref{eq:22}, we have suppressed the second argument, $ \alpha _2 $, in the 3-particle distribution amplitudes for the sake of convenience, and defined the following integrals and operators:
\ba 
	\I [f(u),i,j] := \int du \int d^4x\ e^{i(p+\bar uq)\cdot x} K_i K_j f(u) 
\ea 
\ba 
\I _2 [f(u),i,j] := \int du \int d^4x\ e^{i(p+\bar uq)\cdot x} K_i K_j f(u) x^2
\ea 
\ba 
\I _4 [f(u),i,j] := \int du \int d^4x\ e^{i(p+\bar uq)\cdot x} K_i K_j f(u) x^4
\ea 
\ba 
\IG [f(u)\mathcal F(\alpha _i),i,j] := \int du\int \mathcal D \alpha _i \int d^4x\ e^{i(p+(\alpha _1 + u \alpha _3)q)\cdot x} K_iK_j f(u) \mathcal F (\alpha _i) 
\ea 
\ba 
\IG _2 [f(u)\mathcal F(\alpha _i),i,j] := \int du\int \mathcal D \alpha _i \int d^4x\ e^{i(p+(\alpha _1 + u \alpha _3)q)\cdot x} K_iK_j f(u) \mathcal F (\alpha _i) x^2
\ea 
\ba 
\IG _4 [f(u)\mathcal F(\alpha _i),i,j] := \int du\int \mathcal D \alpha _i \int d^4x\ e^{i(p+(\alpha _1 + u \alpha _3)q)\cdot x} K_iK_j f(u) \mathcal F (\alpha _i) x^4
\ea 
\ba 
	\Hat [n,f(u)] := i^n \int _0^u dv_n \cdots \int _0^{v_3} dv_2 \int _0^{v_2} dv_1\ f(v_1)
\ea 
\ba 
	\Hat _G [n,\mathcal F(\alpha _i)] := (-iu)^n \int _0^{\alpha _3} d\alpha _3^{(n)} \cdots \int _0^{\alpha _3^{(3)}} d\alpha _3^{(2)} \int _0^{\alpha _3^{(2)}} d\alpha _3^{(1)}\ \mathcal F (\alpha _1, 1-\alpha _1 - \alpha _3 ^{(1)}, \alpha _3 ^{(1)})
\ea 
\ba 
	\Der [n, f(u)] := (-i \frac{\del}{\del u})^n f(u)
\ea 
\ba 
	\Der _G [n, \mathcal F(\alpha _i)] := (\frac iu \frac{\del }{\alpha _3}) ^n \mathcal F (\alpha _i)
\ea 
where we have introduced the short-hand notation 
\ba 
	K _i := \frac{K_i(m_Q\sqrt{-x^2})}{(\sqrt{-x^2})^i}, \quad K _j := \frac{K_j (m_{Q'}\sqrt{-x^2})}{(\sqrt{-x^2})^j}
\ea 
As one can see, there are numerous types of integrals appearing in the calculation of the theoretical part of the correlation function. As an example, we present the complete steps of calculations for one type of the integrals, and the remaining ones are presented in Appendix \ref{app:B}.
\prg 
Let us consider the term
\ba 
	\int du \int d^4x\ e^{i(p+\bar uq)\cdot x} \frac{K_i(m_Q \sqrt{-x^2})}{(\sqrt{-x^2})^i} \frac{K_j (m_{Q'} \sqrt{-x^2})}{(\sqrt{-x^2})^j} f(u) 
\ea 
where $ f(u) $ is a generic 2-particle DA of the pseudoscalar meson. Now we use the integral representation of the Bessel function, namely
\ba 
	\frac{K_i (m_Q \sqrt{-x^2})}{(\sqrt{-x^2})^i} = \frac 12 \int _0 ^\infty dt\ \frac{1}{t^{i+1}} e^{-\frac{m_Q}{2}(t - x^2/t)}
\ea 
Then we introduce two new variables $ a $ and $ b $ as $ a := \frac{2m _Q}{t} $ and $ b := \frac{2m _{Q'}}{t'} $. If we now do a Wick rotation, $ x^0 \to i x_E^0 $, and switch to the Euclid spacetime, $ -x^2 \to x _E^2 $, and perform the integration over $ d^4x_E $, we obtain
\ba 
	\frac i4 \frac{16\pi^2}{(2m_Q)^i (2m_{Q'})^j} \int du \int _0^\infty da \int _0^\infty db\ \frac{a^{i-1} b^{j-1}}{(a+b)^2} f(u) e^{-k_E^2/(a+b)} e^{-m_Q^2/a-m_{Q'}^2/b} 
\ea 
where $ k _E := p _E + \bar u q _E $. Now, let us insert the identity $ \int d\rho \ \delta(\rho-a-b) = 1 $, make a scale transformation $ a\to \rho\alpha $, $ b\to \rho\beta $, and perform the integration over $ \beta $ to obtain
\ba 
	\frac i4 \frac{16\pi^2}{(2m_Q)^i (2m_{Q'})^j} \int du \int _0^\infty d\alpha \int d\rho\ e^{-k_E^2/\rho} \rho^{i+j-3} \alpha ^{i-1} (1-\alpha)^{j-1} e^{-(\frac{m_Q^2}{\alpha} + \frac{m_{Q'}^2}{1-\alpha})/\rho} f(u)
\ea 
Then, by writing out
\ba 
	k_E^2 = p_E^2 (1-\bar u) - m_V^2 (\bar u^2 - \bar u) + \bar u (p_E + q_E)^2  
\ea 
and making use of the formula $ \mathfrak B e^{-\alpha p_E^2} = \delta (\frac{1}{M^2} - \alpha) $ to perform the Borel transformations over the variables $ p_E^2 $ and $ (p_E+q_E)^2 $ and integrating over $ u $ and $ \rho $, we get
\ba 
	\frac i4 \frac{16\pi^2}{(2m_Q)^i (2m_{Q'})^j} (M^2)^{i+j} e^{-m_V^2/2M^2} f(\frac 12) \int _0^\infty d\alpha\ \alpha^{i-1} (1-\alpha)^{j-1} e^{-(\frac{m_Q^2}{\alpha} + \frac{m_{Q'}^2}{1-\alpha})/M^2} 
\ea 
Now let
\ba 
	s := \frac{m_Q^2}{\alpha} + \frac{m_{Q'}^2}{1-\alpha}
\ea 
Equation this to $ s_0 $ in order to perform the continuum subtraction, we find the bounds of $ \alpha $. As a result, we obtain
\ba 
	\frac i4 \frac{16\pi^2}{(2m_Q)^i (2m_{Q'})^j} (M^2)^{i+j} e^{-m_V^2/2M^2} f(\frac 12) \int _{\alpha _-}^{\alpha_+} d\alpha\ \alpha^{i-1} (1-\alpha)^{j-1} e^{-(\frac{m_Q^2}{\alpha} + \frac{m_{Q'}^2}{1-\alpha})/M^2}
\ea 
which can be more conveniently rewritten as
\ba 
	& \frac i4 \frac{16\pi^2}{(2m_Q)^i (2m_{Q'})^j} (M^2)^{i+j} e^{-m_V^2/2M^2} f(\frac 12) \int _{(m_Q+m_{Q'})^2}^{s_0} ds \ e^{-s/M^2}\nonumber \\& \times \int d\alpha\ \alpha^{i-1} (1-\alpha)^{j-1} \delta (s - \frac{m_Q^2}{\alpha} - \frac{m _{Q'}^2}{1-\alpha})
\ea
where we have
\ba 
\alpha _\pm := \frac{1}{2s} (s + m_Q^2 - m_{Q'}^2 \pm \sqrt{(s+m_Q^2-m_{Q'}^2)^2 - 4s m_Q^2})
\ea 
and we have defined
\ba
	\frac{1}{M^2} = \frac{1}{M_1^2} + \frac{1}{M_2^2} 
\ea  
Since in our case the mass of the initial and final baryons are practically the same, we take $ M_1^2 = M_2^2 $. The 1/2 inside $ f $ is indeed given by $ u_0 $ which is defined as
\ba 
	u _0 = \frac{M_1^2}{M_1^2 + M_2^2} = \frac 12 
\ea 
Performing similar calculations for the remaining integrals and matching the two representations of the correlation function for the relevant coupling constants, we get the following sum rules: 
\ba 
f_1 &=& \frac{1}{2\lambda _{B_1} \lambda _{B_2}} e^{\frac{m _B^2}{M^2} + \frac{m_V^2}{4M^2}} \Pi ^{\rm theo} _{f_1} \\
f_1 + f_2 &=& \frac{1}{\lambda _{B_1} \lambda _{B_2}} e^{\frac{m _B^2}{M^2} + \frac{m_V^2}{4M^2}} \Pi ^{\rm theo}_{f_1+f_2}
\ea 
\section{Numerical analysis}\label{sec:3}
\begin{figure}
	[H]\centering
	\includegraphics[width=.8\textwidth]{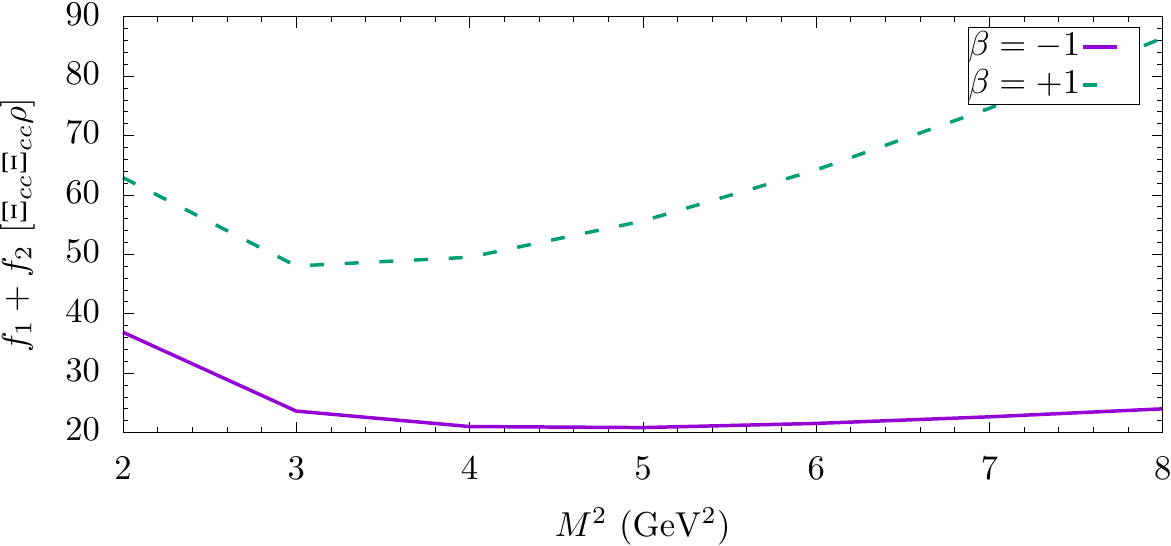}
	\caption{The dependence of the coupling constant $ f_1+f_2 $ for $ \Xi _{cc} \Xi _{cc} \rho $ on $ M^2 $ for the shown values of $ \beta $ and at $ \sqrt{s_0} = 4.6 $ GeV.}
	\label{fig:1}
\end{figure}

\begin{figure}
	[H]\centering
	\includegraphics[width=.8\textwidth]{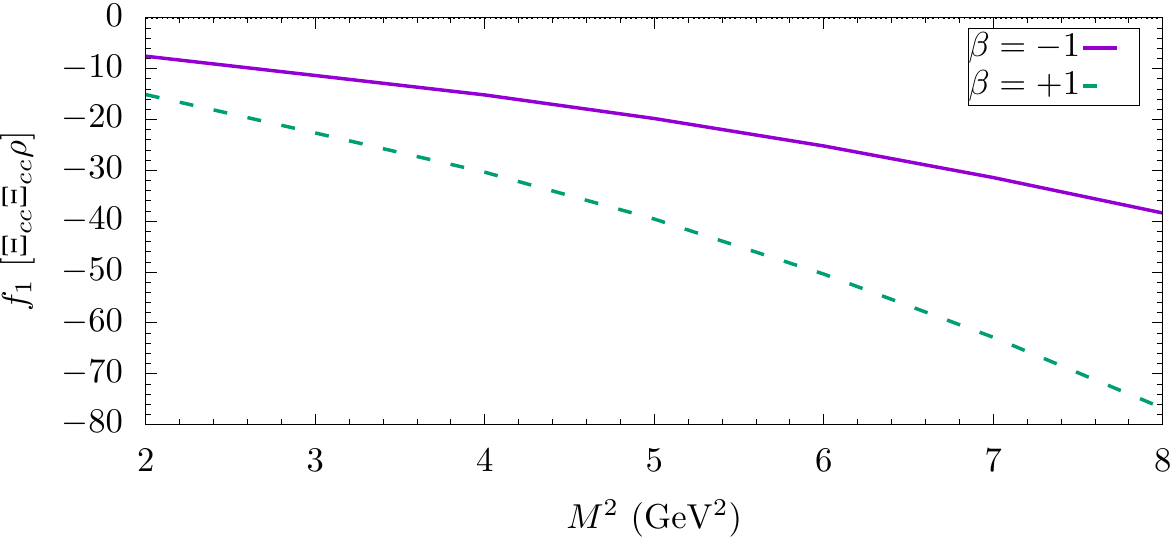}
	\caption{The dependence of the coupling constant $ f_1 $ for $ \Xi _{cc} \Xi _{cc} \rho $ on $ M^2 $ for the shown values of $ \beta $ and at $ \sqrt{s_0} = 4.6 $ GeV.}
	\label{fig:2}
\end{figure}

\begin{figure}
	[H]\centering
	\includegraphics[width=.8\textwidth]{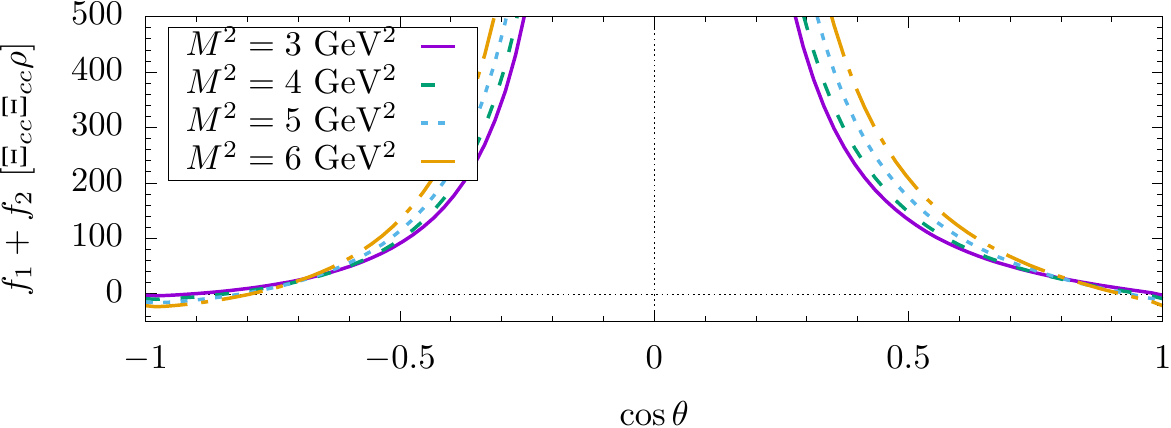}
	\caption{The dependence of the coupling constant $ f_1+f_2 $ for $ \Xi _{cc} \Xi _{cc} \rho $ on $ \cos \theta $ for the shown values of $ M^2 $ and at $ \sqrt{s_0} = 4.6 $ GeV.}
	\label{fig:7}
\end{figure}

\begin{figure}
	[H]\centering
	\includegraphics[width=.8\textwidth]{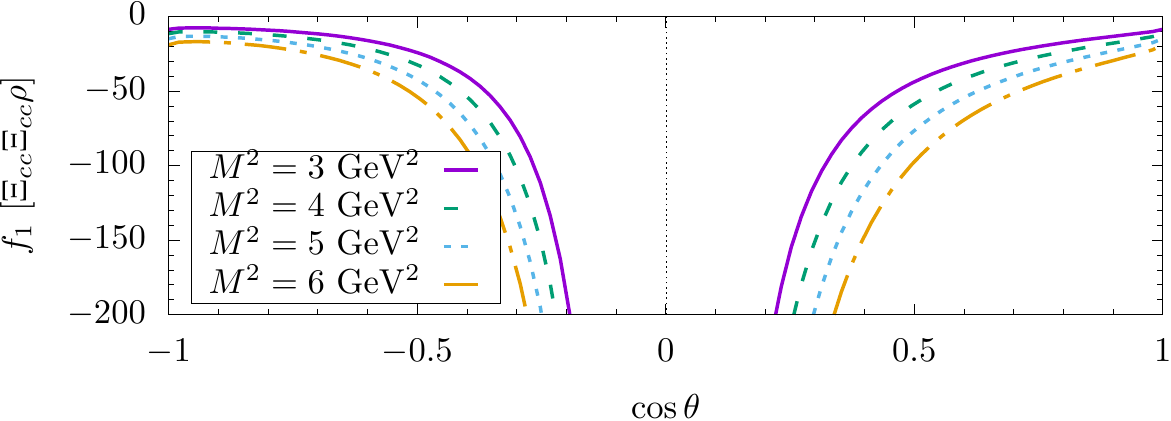}
	\caption{The dependence of the coupling constant $ f_1 $ for $ \Xi _{cc} \Xi _{cc} \rho $ on $ \cos \theta $ for the shown values of $ M^2 $ and at $ \sqrt{s_0} = 4.6 $ GeV.}
	\label{fig:8}
\end{figure}

\begin{figure}
	[H]\centering
	\includegraphics[width=.8\textwidth]{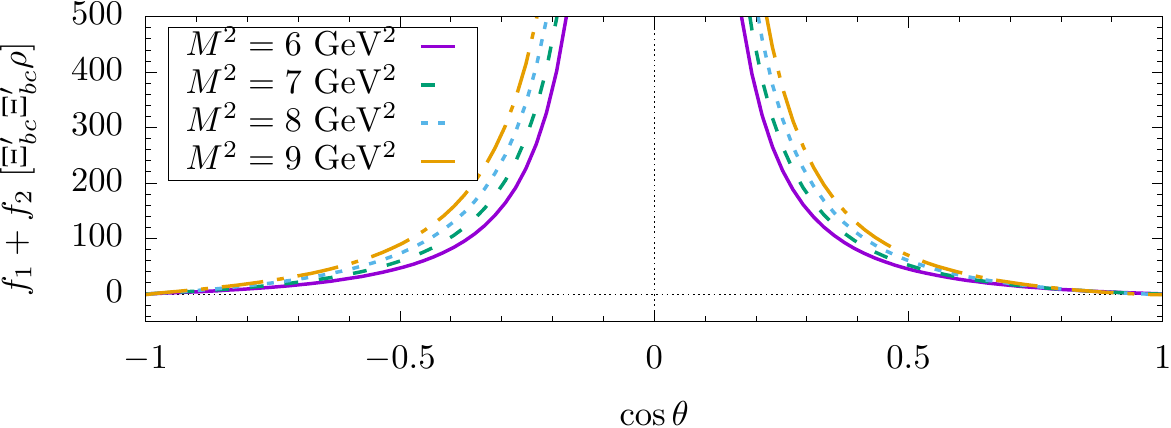}
	\caption{The dependence of the coupling constant $ f_1+f_2 $ for $ \Xi' _{bc} \Xi' _{bc} \rho $ on $ \cos \theta $ for the shown values of $ M^2 $ and at $ \sqrt{s_0} = 7.5 $ GeV.}
	\label{fig:9}
\end{figure}

\begin{figure}
	[H]\centering
	\includegraphics[width=.8\textwidth]{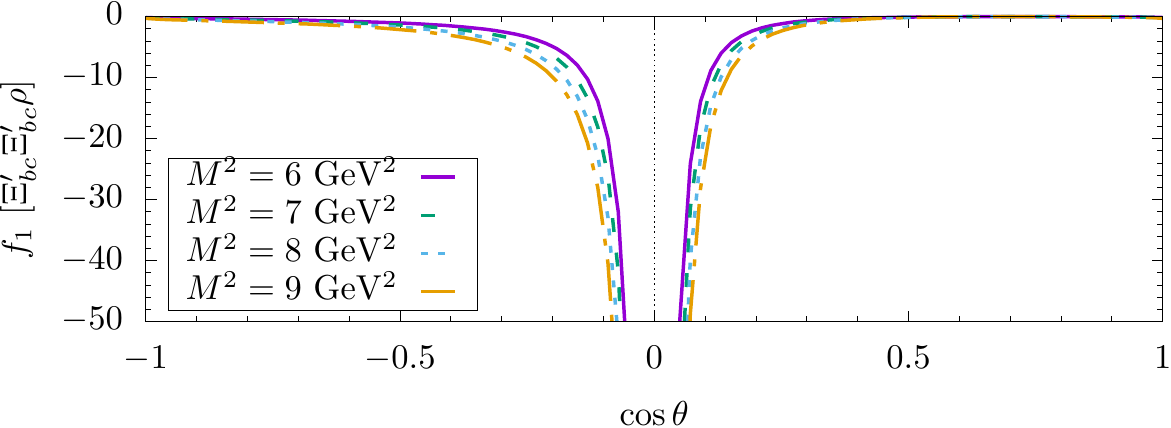}
	\caption{The dependence of the coupling constant $ f_1 $ for $ \Xi' _{bc} \Xi' _{bc} \rho $ on $ \cos \theta $ for the shown values of $ M^2 $ and at $ \sqrt{s_0} = 7.5 $ GeV.}
	\label{fig:10}
\end{figure}

In this section, we numerically analyze the LCSR for the electric-, $ f _1 $, and magnetic-type, $ f _1 + f _2 $, strong coupling constants of the vector meson $ \rho $ and $ K^* $ with the doubly heavy baryons $ \Xi _{cc} $, $ \Xi _{bb} $, $ \Xi _{bc} $, $ \Xi '_{bc} $, $ \Omega _{cc} $, $ \Omega _{bb} $, $ \Omega _{bc} $, and $ \Omega '_{bc} $ by using Package X \cite{ref:package-x}. The LCSR for the coupling constants $ f_1 $ and $ f_1+f_2 $ contain certain input parameters such as quark masses, the masses and decay constants of the vector mesons $ \rho $ and $ K^* $, and the masses and residues of the aforementioned doubly heavy baryons. Some of these parameters are presented in Table \ref{tab:1}. Other input parameters are present in the vector meson DAs of different twists. The complete list of these DAs together with the most recent values of the input parameters are given in Appendix \ref{app:A}. 
{
\renewcommand{\arraystretch}{1.5}
\begin{table}
	[H]\centering 
	\caption{The values of the input parameters entering the sum rules. All the masses and decay constants are in units of GeV.}
	\label{tab:1}
	\begin{tabular}
		{|cc|cc|cc|cc|}
		\hline 
		\hline 
		Parameter 		& Value 	& Parameter 	& Value 	& Parameter 			& Value & Parameter	& Value\\
		\hline 
		$ m_s $ (1 GeV) & 0.137 	& $ m_\rho $	& 0.770  	& $ m_{\Xi_{cc}} $		& 3.72 \cite{ref:onceki-15}	& $ \lambda_{\Xi_{cc}} $&0.16 \cite{ref:onceki-15}\\
		$ m_c $ 		& 1.4  		& $ f_\rho $	& 0.216 	& $ m_{\Xi_{bb}} $		& 9.96 \cite{ref:onceki-15}	& $ \lambda_{\Xi_{bb}} $&0.44 \cite{ref:onceki-15}\\
		$ m_b $ 		& 4.7  		& $ f_\rho^T $ 	& 0.165 	& $ m_{\Xi_{bb}} $		& 6.72 \cite{ref:onceki-15}	& $ \lambda_{\Xi_{bb}} $&0.28 \cite{ref:onceki-15}\\
						&			& $ m_{K^*} $	& 0.892  	& $ m_{\Xi'_{bc}} $		& 6.79 \cite{ref:onceki-15}	& $ \lambda_{\Xi'_{bc}} $&0.30 \cite{ref:onceki-15}\\
						&			& $ f_{K^*} $	& 0.220  	& $ m_{\Omega_{cc}} $	& 3.73 \cite{ref:onceki-15}	& $ \lambda_{\Omega_{cc}} $&0.18 \cite{ref:onceki-15}\\
						&			& $ f_{K^*}^T $	& 0.185  	& $ m_{\Omega_{bb}} $	& 9.97 \cite{ref:onceki-15}	& $ \lambda_{\Omega_{bb}} $&0.45 \cite{ref:onceki-15}\\
						&			&				&			& $ m_{\Omega_{bc}} $	& 6.75 \cite{ref:onceki-15}	& $ \lambda_{\Omega_{bc}} $&0.29 \cite{ref:onceki-15}\\
						&			&				&			& $ m_{\Omega'_{bc}} $	& 6.80 \cite{ref:onceki-15}	& $ \lambda_{\Omega'_{bc}} $&0.31 \cite{ref:onceki-15}\\
		\hline 
		\hline 
	\end{tabular}
\end{table}
}
In addition to the above-mentioned input parameters, the LCSR also includes three auxiliary parameters, namely the Borel mass parameter, $ M^2 $, the continuum threshold, $ s_0 $, and the arbitrary parameter, $ \beta $, which appear in the expression for the interpolating current. Physically measurable quantities should be independent of these parameters. Thus, we need to find the working regions of these auxiliary parameters such that the LCSR is reliable. The lower bound of the Borel mass parameter is obtained by requiring the contributions from the highest-twist terms should be considerably smaller than the contributions from the lowest-twist terms. The upper bound of $ M^2 $ can be determined by demanding that the continuum contribution should not be too large. Meantime, the continuum threshold, $ s _0 $, is obtained by requiring that the two-point sum rules reproduce a 10\% accuracy of the mass of doubly heavy baryons. These conditions lead to the values of $ M^2 $ and $ s _0 $ summarized in Table \ref{tab:2} for the channels considered.
{
\renewcommand{\arraystretch}{1.5}
\begin{table}
	[H]\centering 
	\caption{The working region of the parameters $ M^2 $ and $ s_0 $ for the channels considered in our work.}
	\label{tab:2}
	\begin{tabular}
		{|c|c|c|c|}
		\hline 
		\hline 
		 & Channel & $ M^2 {\rm\ (GeV^2)} $ & $ \sqrt{s_0} {\rm\ (GeV)} $\\
		\hline 
		\multirow{5}{*}{$ SS $} & $ \Xi _{cc} \to \Xi_{cc} \rho $ & $ 3 \leq M^2 \leq 6 $ & 4.6\\
		& $ \Xi _{bb} \to \Xi _{bb} \rho $ & $ 10 \leq M^2 \leq 15 $ & 10.9\\
		& $ \Omega _{bb}\to \Xi _{bb} K^* $ & $ 10 \leq M^2 \leq 15 $ & 10.9 \\
		& $ \Omega _{cc}\to \Xi _{cc} K^* $ & $ 3\leq M^2 \leq 6 $ & 4.6\\
		& $ \Omega _{bc}\to \Xi _{bc} K^* $ & $ 6 \leq M^2 \leq 9 $ & 7.5 \\
		\hline 
		\multirow{2}{*}{$ AA $} & $ \Xi'_{bc}\to \Xi'_{bc}\rho $ & $ 6 \leq M^2 \leq 9 $ & 7.5\\
		& $ \Omega' _{bc}\to \Xi' _{bc} K^* $ & $ 6 \leq M^2 \leq 9 $ & 7.5\\
		\hline 
		\multirow{2}{*}{$ SA $} & $ \Xi'_{bc}\to \Xi_{bc}\rho $ & $ 6 \leq M^2 \leq 9 $ & 7.5\\
		& $ \Omega' _{bc}\to \Xi _{bc} K^* $ & $ 6 \leq M^2 \leq 9 $ & 7.5\\
		\hline
		\hline 
	\end{tabular}
\end{table}
}
Our analysis reveals that the twist-4 term contributions in the aforementioned domains of $ M^2 $ at the indicated values of $ s_0 $ are smaller than 15\% and higher states contribute 30\% at maximum for all the considered channels. As an illustration, we present the $ M^2 $ dependence of $ f_1+f_2 $ and $ f_1 $ for $ \Xi _{cc}\Xi _{cc}\rho $ at fixed values of $ s_0 $ and $ \beta $ in Figs. \ref{fig:1} and \ref{fig:2}. Once we determine the working regions of $ M^2 $ and $ s_0 $, we need to find the working region of the auxiliary parameter, $ \beta $. To this end, we investigate the strong coupling constant $ f_1+f_2 $ as a function of $ \cos \theta $, where we have defined $ \theta $ via $ \beta = \tan \theta $. As an example, we give the dependence of the coupling constants $ f_1+f_2 $ and $ f_1 $ for $ \Xi _{cc}\Xi _{cc}\rho $ and $ \Xi' _{bc} \Xi '_{bc}\rho $ at fixed values of $ M^2 $ and $ s_0 $ in Figs. \ref{fig:7} -- \ref{fig:10}, respectively. 
In these figures, one can see that the coupling constants do not practically change when $ \vert \cos \theta \vert $ varies between 0.6 and 1. Our numerical analysis for the strong coupling constants of doubly heavy baryons with vectors mesons leads to the results presented in Table \ref{tab:3}. The uncertainties are due to the variation of the parameters $ M^2 $, $ s_0 $, and the errors in the values of the input parameters. 
{
\renewcommand{\arraystretch}{1.5}
\begin{table}
	[H]\centering 
	\caption{The numerical values for the strong coupling constants.} 
	\label{tab:3}
	\begin{tabular}
		{|c|c|c|c|}
		\hline 
		\hline 
		 				& Channel 							& $ f_1+f_2 $ 			& $ f_1 $\\
		\hline 
		\multirow{5}{*}{$ SS $} & $ \Xi _{cc}\to \Xi _{cc} \rho $ 		& $ 32.53 \pm 2.30 $	& $ -25.32 \pm 8.49 $\\
								& $ \Xi _{bb}\to \Xi _{bb} \rho $ 		& $ 23.66 \pm 2.89 $	& $ -7.69 \pm 2.31 $ \\
								& $ \Omega _{bb}\to \Xi _{bb} K^* $ 	& $ 22.55 \pm 2.63 $	& $ -8.30 \pm 2.74 $\\
								& $ \Omega _{cc}\to \Xi _{cc} K^* $ 	& $ 28.36 \pm 1.83 $	& $ -23.64 \pm 9.85 $ \\
								& $ \Omega _{bc}\to \Xi _{bc} K^* $ 	& $ 8.23 \pm 0.77 $		& $ -3.74 \pm 1.37 $ \\
		\hline 
		\multirow{2}{*}{$ AA $} & $ \Xi '_{bc}\to \Xi '_{bc}\rho $ 	& $ -37.62 \pm 6.48 $	& $ -0.40 \pm 0.12 $\\
								& $ \Omega ' _{bc}\to \Xi '_{bc} K^* $	& $ -37.98 \pm 6.75 $	& $ -0.40 \pm 0.14 $ \\
		\hline 
		\multirow{2}{*}{$ SA $} & $ \Xi' _{bc}\to \Xi _{bc} \rho $ 	& $ 1.50 \pm 0.31 $ 	& $ -0.97 \pm 0.30 $\\
								& $ \Omega ' _{bc}\to \Xi _{bc} K^* $ 	& $ 2.20 \pm 0.40 $		& $ -1.00 \pm 0.36 $ \\
		\hline 
		\hline 
	\end{tabular}
\end{table}
}
From Table \ref{tab:3}, we deduce the following conclusions:
\begin{enumerate}
	[(a)]
	\item The $ SU(3)$ symmetry for the $ SS$ and $ AA$ cases works very well. The violation of the $ SU(3) $ symmetry is about 10\% at maximum.
	\item In the SA case, SU(3) symmetry works very well for electric strong coupling, $ f_1 $, but considerably violated for the coupling $ f_1+f_2 $ (about 50\%). 
\end{enumerate}
\section{Conclusion}\label{sec:4}
The experimental discovery of the $ \Xi _{cc} $ baryon opened a new research area in theoretical studies for understanding the properties of doubly heavy baryons by analyzing their weak, electromagnetic, and strong decays. In the present work, within the framework of the LCSR method, we estimate the electric and magnetic couplings of light vector mesons $ \rho $ and $ K^* $ with doubly heavy baryons of spin 1/2. In our analysis, we have used the general form of the interpolating currents in symmetric and antisymmetric forms with respect to the exchange of two heavy quarks. We obtained that in the case of symmetric-antisymmetric ($ SA $) currents, couplings are affirmed to be smaller than the $ SS $ and $ AA $ cases. This circumstance can be explained by the fact that in the $ SA $ case, there are strong cancellations between the leading terms. Moreover, we obtained that our results for the considered couplings constants for the cases of $ SS $ and $ AA $ are in a good agreement with the $ SU(3) $ symmetry results and its violation is about 10\% at maximum. For the $ SA $ case, for the electric coupling, the $ SU(3) $ symmetry violation is small (about 5\%), but for magnetic coupling, its violation is about 50\%. 
\prg 
Our final remark is that these results can be helpful in studies of the properties of doubly heavy baryons in experiments conducted at LHCb. 
%\prg 
%In this paper, we have investigated the electric- and magnetic-type strong coupling constants of the vector mesons $ \rho $ and $ K^* $ with the doubly heavy baryons $ \Xi _{QQ'} $, $ \Xi '_{QQ'} $, $ \Omega _{QQ'} $, and $ \Omega '_{QQ'} $, where $ Q, Q' = \{b,c \} $ within the framework of light-cone QCD sum rules. The numerical values of the electric and magnetic couplings have been obtained.
\appendix
\section{Distribution amplitudes for vector mesons}\label{app:A}
In this section, we collect the matrix elements $ \langle V(q,\varepsilon) \vert \bar q(x) \Gamma _i q(0) \vert 0 \rangle $ and $ \langle V(q,\varepsilon) \vert \bar q(x) \Gamma _i G_{\mu\nu} q(0) \vert 0 \rangle $ and the relevant distribution amplitudes for vector mesons together with the most recent values for the DA parameters involved \cite{ref:ball-1, ref:ball-2, ref:ball-3, ref:ball-4}.
\prg
Up to twist-4 accuracy, the matrix elements $ \langle V(q,\varepsilon) \vert \bar q(x) \Gamma q(0) \vert 0 \rangle $ and $ \langle V(q,\varepsilon) \vert \bar q(x) \Gamma G_{\mu\nu} q(0) \vert 0 \rangle $ are given as follows:
\ba
  \langle V(q,\varepsilon) \vert \bar q_1 (x) \gamma_\mu q_2(0) \vert 0 \rangle &=& f_V m_V (
    \frac{\varepsilon^\lambda \cdot x}{q\cdot x} q_\mu \int_0^1 du\ e^{i\bar u q\cdot x} (\phi_2^\parallel (u) + \frac{m_V^2x^2}{16} \phi_4^\parallel(u)) 
    \ar (\varepsilon_\mu^\lambda - q_\mu \frac{\varepsilon^\lambda \cdot x}{q\cdot x}) \int_0^1 du\ e^{i\bar u q\cdot x} \phi_3^\perp (u)
    \ek \frac12 x_\mu \frac{\varepsilon^\lambda \cdot x}{(q\cdot x)^2} m_V^2 \int _0^1du\ e^{i\bar uq\cdot x} (\psi _4^\parallel(u) + \phi_2^\parallel (u) - 2 \phi_3^\perp (u))  
)
\ea
\ba
  \langle V(q,\varepsilon) \vert \bar q_1(x) \gamma_\mu \gamma_5 q_2(0) \vert 0 \rangle = -\frac14 \epsilon _{\mu}^{\nu\alpha\beta} \varepsilon^\lambda_\nu q_\alpha x_\beta f_V m_V \int_0^1 du\ \psi_3^\perp (u)
\ea
\ba
  \langle V(q,\varepsilon) \vert \bar q_1(x) \sigma_{\mu\nu} q_2 (0) \vert 0 \rangle &=& -if_V^T (
    (\varepsilon^\lambda_\mu q_\nu - \varepsilon^\lambda_\nu q_\mu) \int_0^1du\ e^{i\bar u q \cdot x} \phi_2^\perp(u) + \frac{m_V^2x^2}{16} \phi_4^\perp(u) 
    \ar \frac{\varepsilon \cdot x}{(q\cdot x)^2} (q_\mu x_\nu - q_\nu x_\mu) \int_0^1du \ e^{i\bar u q \cdot x} (\phi_3^\parallel (u) - \frac 12 \phi_2^\perp (u) - \frac 12 \psi_4^\perp(u))
    \ar \frac 12 (\varepsilon^\lambda_\mu x_\nu - \varepsilon^\lambda_\nu x_\mu) \frac{m_V^2}{q\cdot x} \int_0^1 du\ e^{i\bar u q\cdot x} (\psi_4^\perp(u) - \phi_2^\perp(u))
)
\ea
\ba
  \langle V(q,\varepsilon) \vert \bar q_1(x) \sigma_{\alpha\beta} g_s G_{\mu\nu} (ux) q_2(0) \vert 0 \rangle &=& f_V^T m_V^2 \frac{\varepsilon^\lambda \cdot x}{2q\cdot x} (q_\alpha q_\mu g_{\beta\nu}^\perp - q_\beta q_\mu g_{\alpha\nu}^\perp - q_\alpha q_\nu g_{\beta\mu}^\perp + q_\beta q_\nu g_{\alpha\mu}^\perp) 
  \carp \int \mathcal D \alpha_i \ e^{i(\alpha_1 + u\alpha_3)q\cdot x} \mathcal T(\alpha _i) 
  \ar f_V^T m_V^2 (q_\alpha \varepsilon_\mu^\lambda g_{\beta\nu}^\perp - q_\alpha \varepsilon_\nu^\lambda g_{\beta\mu}^\perp + q_\beta \varepsilon^\lambda_nu g_{\alpha\mu}^\perp) 
  \carp \int \mathcal D\alpha_i \ e^{i(\alpha_1 + u \alpha_3)q\cdot x} \mathcal T_1^{(4)}(\alpha _i)
  \ar f_V^T m_V^2 (q_\mu \varepsilon_\alpha^\lambda g_{\beta\nu}^\perp - q_\mu \varepsilon_\beta^\lambda g_{\alpha\nu}^\perp - q_\nu \varepsilon_\alpha^\lambda g_{\beta\mu}^\perp + q_\nu \varepsilon_\beta^\lambda g_{\alpha\mu}^\perp) 
  \carp \int \mathcal D \alpha_i \ e^{i(\alpha_1 + u \alpha_3)q\cdot x} \mathcal T_2^{(4)}(\alpha _i)
  \ar \frac{f_V^T m_V^2}{q\cdot x} (q_\alpha q_\mu \varepsilon^\lambda _\beta x_\nu - q_\beta q_\mu \varepsilon_\alpha^\lambda x_\nu - q_\alpha q_\nu \varepsilon^\lambda_\beta x_\mu + q_\beta q_\nu \varepsilon_\alpha^\lambda x_\mu) 
  \carp \int \mathcal D \alpha_i \ e^{i(\alpha_1 + u \alpha_3)q\cdot x} \mathcal T_3^{(4)}(\alpha _i)
  \ar \frac{f_V^T m_V^2}{q\cdot x} (q_\alpha q_\mu \varepsilon_\nu^\lambda x_\beta - q_\beta q_\mu \varepsilon_\nu^\lambda x_\alpha - q_\alpha q_\nu \varepsilon^\lambda_\mu x_\beta + q_\beta q_\nu \varepsilon^\lambda_\mu x_\alpha) 
  \carp \int \mathcal D \alpha_i \ e^{i(\alpha_1 + u \alpha_3)q\cdot x} \mathcal T_4^{(4)}(\alpha _i)
\ea
\ba
  \langle V(q,\varepsilon) \vert \bar q_1(x) g_s G_{\mu\nu} (ux) q_2(0) \vert 0 \rangle = -i f_V^T m_V (\varepsilon^\lambda_\mu q_\nu - \varepsilon^\lambda_\nu q_\mu) \int \mathcal D \alpha _i \ e^{i(\alpha_1 + u \alpha_3)q\cdot x} \mathcal S(\alpha _i)
\ea
\ba
  \langle V(q,\lambda) \vert \bar q_1(x) g_s \tilde G_{\mu\nu} (ux)\gamma_5 q_2(0) \vert 0 \rangle = -i f_V^T m_V (\varepsilon_\mu^\lambda q_\nu - \varepsilon^\lambda_\nu q_\mu) \int\mathcal D \alpha _i \ e^{i(\alpha_1 + u\alpha_3) q\cdot x} \tilde{\mathcal S} (\alpha _i) 
\ea
\ba
  \langle V(q,\varepsilon) \vert \bar q_1(x) g_s \tilde G_{\mu\nu} (ux) \gamma_\alpha \gamma_5 q_2(0) \vert 0 \rangle = f_V m_V q_\alpha (\varepsilon^\lambda_\mu q_\nu - \varepsilon^\lambda_nu q_\mu) \int\mathcal D\alpha_i\ e^{i(\alpha_1 + u \alpha_3)q\cdot x} \mathcal A(\alpha_i)
\ea
\ba
  \langle V(q,\varepsilon) \vert \bar q_1(x) g_s G_{\mu\nu} (ux) i\gamma_\alpha q_2(0) \vert 0 \rangle = f_V m_V q_\alpha (\varepsilon^\lambda_\mu q_\nu - \varepsilon^\lambda_\nu q_\mu) \int \mathcal D \alpha _i \ e^{i(\alpha_1 + u \alpha_3)q\cdot x} \mathcal V(\alpha_i)
\ea
where $ \tilde G _{\mu\nu} = \frac12 \epsilon_{\mu\nu\alpha\beta} G^{\alpha\beta} $ is the dual gluon field strength tensor and $ \int \mathcal D\alpha_i = \int d\alpha_1 \ d\alpha_2 d\alpha_3 \delta(1-\alpha_1 - \alpha_2 - \alpha_3) $. Now we list the DAs.
\prg 
\textit{2-particle twist-2 DAs:}
\ba
\phiiiparu =
	6 \baru (1 + \aipar C_{1}^{3/2}(\xi) + \aiipar C_{2}^{3/2}(\xi)) u
\ea
\ba
\phiiiperu = 
	6 \baru (1 + \aiper C_{1}^{3/2}(\xi) + \aiiper C_{2}^{3/2}(\xi)) u
\ea
\textit{2-particle twist-3 DAs:}
\ba
\phiiiiparu &=& 
  3 \xi^2 
  + (3 \aiper \xi (-1 + 3 \xi^2))/2 
  + ((15 \kappaiiiper)/2 - (3 \lambdaiiiper)/4) \xi (-3 + 5 \xi^2) 
  \ar (3 \aiiper \xi^2 (-3 + 5 \xi^2))/2 
  + (5 \omegaiiiper (3 - 30 \xi^2 + 35 \xi^4))/8 
  - (3 \fV (\mqi - \mqii) \xi (2 + 9 \aipar \xi \carp 2 \aiipar (11 - 30 \baru u) + (1 + 6 \aiipar + 3 \aipar) \ln(\baru) + (1 + 6 \aiipar - 3 \aipar) \ln(u)))/(2 \fVT \mV) 
  \ar (3 \fV (\mqi + \mqii) (1 + 8 \aipar \xi + 3 \aiipar (7 - 30 \baru u) + (1 + 6 \aiipar + 3 \aipar) \xi \ln(\baru) - (1 + 6 \aiipar \ek 3 \aipar) \xi \ln(u)))/(2 \fVT \mV)
\ea
\ba
\psiiiiparu &=&
	6 \baru (1 + C_{1}^{3/2}(\xi) (\aiper/3 + (5 \kappaiiiper)/3) - (C_{3}^{3/2}(\xi) \lambdaiiiper)/20 + C_{2}^{3/2}(\xi) (\aiiper/6 + (5 \omegaiiiper)/18)) 
  \carp u - (3 \fV (\mqi - \mqii) (\baru (9 \aipar + 10 \aiipar \xi) u + (1 + 6 \aiipar + 3 \aipar) \baru \ln(\baru) - (1 + 6 \aiipar - 3 \aipar) 
  \carp u \ln(u)))/(\fVT \mV) + (3 \fV (\mqi + \mqii) (\baru u (1 + 2 \aipar \xi + 3 \aiipar (7 - 5 \baru u)) + (1 + 6 \aiipar 
  \ar 3 \aipar) \baru \ln(\baru) + (1 + 6 \aiipar - 3 \aipar) u \ln(u)))/(\fVT \mV)
\ea
\ba
\psiiiiperu &=&
	6 \baru (1 + C_{1}^{3/2}(\xi) (\aipar/3 + (20 \kappaiiipar)/9) + C_{3}^{3/2}(\xi) (-\lambdaiiipar/8 + \lambdaiiipartilde/4) + C_{2}^{3/2}(\xi) (\aiipar/6 
  \ar (5 \omegaiiipar)/12 - (5 \omegaiiipartilde)/24 + (10 \zetaiiipar)/9)) u - (6 \fVT (\mqi - \mqii) (\baru (9 \aiper + 10 \aiiper \xi) u + (1 
  \ar 6 \aiiper + 3 \aiper) \baru \ln(\baru) - (1 + 6 \aiiper - 3 \aiper) u \ln(u)))/(\fV \mV) + (6 \fVT (\mqi + \mqii) (\baru u 
  \carp (2 + 3 \aiper \xi + 2 \aiiper (11 - 10 \baru u)) + (1 + 6 \aiiper + 3 \aiper) \baru \ln(\baru) + (1 + 6 \aiiper - 3 \aiper) 
  \carp u \ln(u)))/(\fV \mV)
\ea
\ba
\phiiiiperu &=&
	(3 \aipar \xi^3)/2 + (3 (1 + \xi^2))/4 + (5 \kappaiiipar - (15 \lambdaiiipar)/16 + (15 \lambdaiiipartilde)/8) \xi (-3 + 5 \xi^2) 
  \ar ((9 \aiipar)/112 + (15 \omegaiiipar)/32 - (15 \omegaiiipartilde)/64) (3 - 30 \xi^2 + 35 \xi^4) + (-1 + 3 \xi^2) ((3 \aiipar)/7 
  \ar 5 \zetaiiipar) - (3 \fVT (\mqi - \mqii) (2 \xi + 2 \aiiper \xi (11 - 20 \baru u) + 9 \aiper (1 - 2 \baru u) + (1 + 6 \aiiper 
  \ar 3 \aiper) \ln(\baru) - (1 + 6 \aiiper - 3 \aiper) \ln(u)))/(2 \fV \mV) + (3 \fVT (\mqi + \mqii) (2 + 9 \aiper \xi 
  \ar 2 \aiiper (11 - 30 \baru u) + (1 + 6 \aiiper + 3 \aiper) \ln(\baru) 
  \ar (1 + 6 \aiiper - 3 \aiper) \ln(u)))/(2 \fV \mV)
\ea
\textit{2-particle twist-4 DAs:}
\ba
\psiivparu &=&
	1 + C_{3}^{1/2}(\xi) ((-9 \aipar)/5 - (20 \kappaiiipar)/3 - (16 \kappaivpar)/3) + C_{1}^{1/2}(\xi) ((9 \aipar)/5 + 12 \kappaivpar) 
	\ar C_{3}^{1/2}(\xi) (-5 \thetaiipar + 10 \thetaipar) + (6 \fVT (\mqi - \mqii) (\xi + (\aiper (-1 + 3 \xi^2))/2 
	\ar (5 \kappaiiiper (-1 + 3 \xi^2))/2 + (\aiiper \xi (-3 + 5 \xi^2))/2 + (5 \omegaiiiper \xi (-3 + 5 \xi^2))/6 
	\ek (\lambdaiiiper (3 - 30 \xi^2 + 35 \xi^4))/16))/(\fV \mV) + C_{4}^{1/2}(\xi) ((-27 \aiipar)/28 - (15 \omegaiiipar)/8 
	\ek (15 \omegaiiipartilde)/16 + (5 \zetaiiipar)/4) + C_{2}^{1/2}(\xi) (-1 - (2 \aiipar)/7 + (40 \zetaiiipar)/3) 
	\ek (20 C_{2}^{1/2}(\xi) \zetaivpar)/3
\ea
\ba
\phiivparu &=&
	(6 \baru \fVT (\mqi - \mqii) (-(C_{1}^{3/2}(\xi) ((82 \aiper)/5 + 10 \kappaiiiper)) + C_{3}^{3/2}(\xi) ((2 \aiper)/5 + (7 \lambdaiiiper)/54) 
	\ar (2 C_{5}^{3/2}(\xi) \lambdaiiiper)/135 + C_{4}^{3/2}(\xi) (-2/315 + \aiiper/5 - \omegaiiiper/21) + 20 C_{2}^{3/2}(\xi) (10/189 
	\ar \aiiper/3 - \omegaiiiper/21)) u)/(\fV \mV) + (6 \baru \fVT (\mqi + \mqii) (2 (3 + 16 \aiiper) + (10 C_{1}^{3/2}(\xi) (-\aiper 
	\ar \kappaiiiper))/3 - (C_{3}^{3/2}(\xi) \lambdaiiiper)/10 + C_{2}^{3/2}(\xi) (-\aiiper + (5 \omegaiiiper)/9)) u)/(\fV \mV) 
	\ar 30 \baru^2 (C_{1}^{5/2}(\xi) ((17 \aipar)/50 - \lambdaiiipar/5 + (2 \lambdaiiipartilde)/5) + (C_{2}^{5/2}(\xi) ((9 \aiipar)/7 + (7 \omegaiiipar)/6 
	\ek (3 \omegaiiipartilde)/4 + \zetaiiipar/9))/10 + (4 (1 + \aiipar/21 + (10 \zetaiiipar)/9))/5) u^2 + 30 \baru^2 (C_{1}^{5/2}(\xi) ((2 \thetaiipar)/3 
	\ek (8 \thetaipar)/15) + (20 \zetaivpar)/9) u^2 + (\fVT (\mqi - \mqii) ((-23 - 108 \aiiper - 54 \aiper + 5 u^2) \ln(\baru) 
	\ek (-23 - 108 \aiiper + 54 \aiper + 5 \baru^2) \ln(u)))/(\fV \mV) + (24 \fVT (\mqi + \mqii) ((1 + 6 \aiiper 
	\ar 3 \aiper) \baru^2 \ln(\baru) + (1 + 6 \aiiper - 3 \aiper) u^2 \ln(u)))/(\fV \mV) + 4 (\aipar - (40 \kappaiiipar)/3) ((11 
	\ek 3 \xi^2)/8 - (2 - \baru) \baru^3 \ln(\baru) + (2 - u) u^3 \ln(u)) + 80 \psiiipar ((11 
	- 3 \xi^2)/8 
	\ek (2 - \baru) \baru^3 \ln(\baru) + (2 - u) u^3 \ln(u)) - 80 \omegaivpartilde ((\baru (21 - 13 \xi^2) u)/8 + \baru^3 (10 - 15 \baru 
	\ar 6 \baru^2) \ln(\baru) + u^3 (10 - 15 u + 6 u^2) \ln(u)) + 2 (-2 \aiipar + 3 \omegaiiipar - (14 \zetaiiipar)/3) ((\baru (21 
	\ek 13 \xi^2) u)/8 + \baru^3 (10 - 15 \baru + 6 \baru^2) \ln(\baru) + u^3 (10 - 15 u + 6 u^2) \ln(u))
\ea
\ba
\psiivperu &=&
	1 + C_{1}^{1/2}(\xi) ((-3 \aiper)/5 + 12 \kappaivper) + (C_{5}^{1/2}(\xi) \lambdaiiiper)/3 + C_{4}^{1/2}(\xi) ((-3 \aiiper)/7 
	\ek (5 \omegaiiiper)/4) + C_{3}^{1/2}(\xi) ((3 \aiper)/5 - 5 \kappaiiiper - 12 \kappaivper - \lambdaiiiper/3 + 5 ((-\thetaiiper - \thetaiipertilde)/2 + \thetaiper 
	\ar \thetaipertilde)) + (\fV (\mqi + \mqii) (3 (1 + 6 \aiipar) + 3 \aipar C_{1}^{1/2}(\xi) + 5 C_{3}^{1/2}(\xi) (4 \kappaiiipar - (3 \lambdaiiipar)/4 
	\ar (3 \lambdaiiipartilde)/2) + (15 C_{4}^{1/2}(\xi) (2 \omegaiiipar - \omegaiiipartilde))/4 + 5 C_{2}^{1/2}(\xi) (-3 \aiipar + 4 \zetaiiipar)))/(\fVT \mV) 
	\ar C_{2}^{1/2}(\xi) (-1 + (3 \aiiper)/7 - 10 (\zetaivper + \zetaivpertilde)) - (6 \baru \fV (\mqi - \mqii) (9 \aipar + 10 \aiipar \xi)
	\carp u)/(\fVT \mV) + (6 \fV (\mqi - \mqii) (-((1 + 6 \aiipar + 3 \aipar) \baru \ln(\baru)) + (1 + 6 \aiipar - 3 \aipar) 
	\carp u \ln(u)))/(\fVT \mV) + (6 \fV (\mqi + \mqii) ((1 + 6 \aiipar + 3 \aipar) \baru \ln(\baru) + (1 + 6 \aiipar - 3 \aipar)
	\carp u \ln(u)))/(\fVT \mV)
\ea
\ba
\phiivperu &=&
	30 \baru^2 (2/5 + (4 \aiiper)/35 - (4 C_{3}^{5/2}(\xi) \lambdaiiiper)/1575 + C_{2}^{5/2}(\xi) ((3 \aiiper)/35 + \omegaiiiper/60) 
	\ar C_{1}^{5/2}(\xi) ((3 \aiper)/25 + \kappaiiiper/3 - \lambdaiiiper/45 + (7 \thetaiiper)/30 - (3 \thetaiipertilde)/20 - \thetaiper/15 + \thetaipertilde/5) 
	\ar (4 \zetaivper)/3 - (8 \zetaivpertilde)/3) u^2 + (-\aiper + 5 \kappaiiiper - 20 \phiiipertilde) ((\baru \xi (-11 + 3 \xi^2) u)/2 + 4 (2 - \baru) 
	\carp \baru^3 \ln(\baru) - 4 (2 - u) u^3 \ln(u)) + ((-36 \aiiper)/11 - (252 \avqi)/55 - (140 \avqiii)/11 
	\ar 2 \omegaiiiper) (-(\baru (-21 + 13 \xi^2) u)/8 + \baru^3 (10 - 15 \baru + 6 \baru^2) \ln(\baru) + u^3 (10 - 15 u + 6 u^2)
	\carp \ln(u))
\ea
\textit{3-particle twist-3 DAs:}
\ba
\SS &=&
	30 \alphaiii^2 (((-3 (\alphai^2 + (1 - \alphai - \alphaiii)^2))/2 + (1 - \alphaiii) \alphaiii) \psiiiper + (-6 \alphai (1 - \alphai - \alphaiii) 
	\ar (1 - \alphaiii) \alphaiii) \psiiper + (1 - \alphaiii) \psioper - (-1 + 2 \alphai + \alphaiii) (((-3 + 5 \alphaiii) \thetaiiper)/2 + \alphaiii \thetaiper 
	\ar \thetaoper))
\ea
\ba
\SSTilde &=&
	30 \alphaiii^2 (((-3 (\alphai^2 + (1 - \alphai - \alphaiii)^2))/2 + (1 - \alphaiii) \alphaiii) \psiiipertilde + (-6 \alphai (1 - \alphai - \alphaiii) 
	\ar (1 - \alphaiii) \alphaiii) \psiipertilde + (1 - \alphaiii) \psiopertilde - (\alphai - \alphaiii) (((-3 + 5 \alphaiii) \thetaiipertilde)/2 + \alphaiii \thetaipertilde 
	\ar \thetaopertilde))
\ea
\ba
\VV =
	360 \alphai (1 - \alphai - \alphaiii) \alphaiii^2 (\kappaiiipar + ((-3 + 7 \alphaiii) \lambdaiiipar)/2 + (-1 + 2 \alphai + \alphaiii) \omegaiiipar)
\ea
\ba
\AA =
	360 \alphai (1 - \alphai - \alphaiii) \alphaiii^2 ((-1 + 2 \alphai + \alphaiii) \lambdaiiipartilde + ((-3 + 7 \alphaiii) \omegaiiipartilde)/2 + \zetaiiipar)
\ea
\ba
\TT =
	360 \alphai (1 - \alphai - \alphaiii) \alphaiii^2 (\kappaiiiper + ((-3 + 7 \alphaiii) \lambdaiiiper)/2 + (-1 + 2 \alphai + \alphaiii) \omegaiiiper)
\ea
\textit{3-particle twist-4 DAs:}
\ba
\TTi =
	120 \alphai (1 - \alphai - \alphaiii) \alphaiii ((-1 + 2 \alphai + \alphaiii) \phiiper + \phioper + (-1 + 3 \alphaiii) \phiiiper)
\ea
\ba
\TTii &=&
	-30 \alphaiii^2 (-((-1 + 2 \alphai + \alphaiii) (((-3 + 5 \alphaiii) \psiiipertilde)/2 + \alphaiii \psiipertilde + \psiopertilde)) + ((-3 (\alphai^2 
	\ar (1 - \alphai - \alphaiii)^2))/2 + (1 - \alphaiii) \alphaiii) \thetaiipertilde + (-6 \alphai (1 - \alphai - \alphaiii) + (1 - \alphaiii) \alphaiii) \thetaipertilde 
	\ar (1 - \alphaiii) \thetaopertilde)
\ea
\ba
\TTiii =
	-120 \alphai (1 - \alphai - \alphaiii) \alphaiii ((-1 + 3 \alphaiii) \phiiipertilde + (-1 + 2 \alphai + \alphaiii) \phiipertilde + \phiopertilde)
\ea
\ba
\TTiv &=&
	30 \alphaiii^2 (-((-1 + 2 \alphai + \alphaiii) (((-3 + 5 \alphaiii) \psiiiper)/2 + \alphaiii \psiiper + \psioper)) + ((-3 (\alphai^2 
	\ar (1 - \alphai - \alphaiii)^2))/2 + (1 - \alphaiii) \alphaiii) \thetaiiper + (-6 \alphai (1 - \alphai - \alphaiii) + (1 - \alphaiii) \alphaiii) \thetaiper 
	\ar (1 - \alphaiii) \thetaoper)
\ea
where we have replaced $ \alpha_2 = 1-\alpha_1-\alpha_3 $ before the integration and we take $ \xi = \bar u $ since the second quark is at the point $ x = 0 $. The $ C_n^m(x) $ are the Gegenbauer polynomials. The $ q _1 $ and $ q_2 $ indicate the quark components of the vector meson. The $ \rho $ meson has both light quarks, hence $ \mqi = \mqii = 0 $. The $ K^* $ meson has one strange quark and one light quark, thus $ \mqi = m_s $ but $ \mqii = 0 $. The derived DA parameters are given as follows:
\ba
  \psioper = \zetaivper
\ea
\ba
  \psiopertilde = \zetaivpertilde
\ea
\ba
  \thetaoper = -(1/6) \kappaiiiper - (1/3) \kappaivper
\ea
\ba
  \thetaopertilde = -(1/6) \kappaiiiper + (1/3) \kappaivper
\ea
\ba
  \phioper = (1/6) \kappaiiiper + (1/3) \kappaivper
\ea
\ba
  \phiopertilde = (1/6) \kappaiiiper - (1/3) \kappaivper
\ea
\ba
  \phiiper = (9/44) \aiiper + (1/8) \omegaiiiper + (63/220) \avqi - (119/44) \avqiii
\ea
\ba
  \phiipertilde = -(9/44) \aiiper + (1/8) \omegaiiiper - (63/220) \avqi - (35/44) \avqiii
\ea
\ba
  \psiiper = (3/44) \aiiper + (1/12) \omegaiiiper + (49/110) \avqi - (7/22) \avqiii + (7/3) \avqv
\ea
\ba
  \psiipertilde = -(3/44) \aiiper + (1/12) \omegaiiiper - (49/110) \avqi + (7/22) \avqiii + (7/3) \avqv
\ea
\ba
  \psiiiper = -(3/22) \aiiper - (1/12) \omegaiiiper + (28/55) \avqi + (7/11) \avqiii + (14/3) \avqv
\ea
\ba
  \psiiipertilde = (3/22) \aiiper - (1/12) \omegaiiiper - (28/55) \avqi - (7/11) \avqiii + (14/3) \avqv
\ea
\ba
  \thetaipar = -(7/10) \aipar \zetaivpar
\ea
\ba
  \thetaiipar = (7/5) \aipar \zetaivpar
\ea
\ba
  \psiiipar = -(7/20) \aipar \zetaivpar
\ea
\ba
  \thetaiper = -(21/10) \zetaivper \aiper
\ea
\ba
  \thetaipertilde = (21/10) \zetaivper \aiper
\ea
\ba
  \thetaiiper = (21/5) \zetaivper \aiper
\ea
\ba
  \thetaiipertilde = -(21/5) \zetaivper \aiper
\ea
\ba
  \phiiipertilde = -(21/20) \zetaivper \aiper
\ea
\ba
  \avqi = -(10/3) \zetaivper
\ea
\ba
  \avqiii = -\zetaivper
\ea
\ba
  \avqv = 0
\ea
The numerical values for the DA parameters are given in Table \ref{tab:a-1}.
\def\birim#1{{\rm #1}}
{\renewcommand\arraystretch{1.5}
	\begin{table}[H]\centering
		\caption{The numerical values for the parameters in the DAs for vector mesons $ \rho $ and $ K^* $. The renormalization scale is $ \mu = 1 \ \birim{GeV} $.}\label{tab:a-1}
		\begin{tabular}{|c|c|c|c|c|c|c|c|c|c|}
			\hline 
			\hline 
			 & $ f_V $ [GeV] & $ f_V^T $ [GeV] & $ m_V $ [GeV] & $ a_1^\parallel $ & $ a_1^\perp $ & $ a_2^\parallel $ & $ a_2^\perp $ & $ \zeta_3^\parallel $ & $ \tilde \lambda _3 ^\parallel $ \\
			\hline 
			$ \rho $ & 0.216 & 0.165 & 0.770 & 0 & 0 & 0.15 & 0.14& 0.03& 0\\
			$ K^* $ & 0.220 & 0.185 & 0.892 & 0.03 & 0.04 & 0.11 &0.1& 0.023&0.035 \\
			\hline 
			\hline
		\end{tabular}\vspace{5mm}
		\begin{tabular}{|c|c|c|c|c|c|c|c|c|c|c|}
			\hline 
			\hline 
			 & $ \tilde \omega_3 ^\parallel $ & $ \kappa_3^\parallel $ & $ \omega_3^\parallel $ & $ \lambda _3 ^\parallel $ & $ \kappa _3^\perp $ & $ \omega_3^\perp $ & $ \lambda_3^\perp $ & $ \zeta_4^\parallel $ & $ \tilde\omega_4^\parallel $ & $ \zeta_4^\perp $\\
			\hline 
			$ \rho $ & --0.09 & 0 & 0.15 & 0 & 0 & 0.55 & 0 & 0.07 & --0.03 & --0.08\\
			$ K^* $ & --0.07 & 0 & 0.1 & --0.008 & 0.003 & 0.3 & --0.025 & 0.02 & --0.02 & --0-.05 \\
			\hline 
			\hline
		\end{tabular}\vspace{5mm}
		\begin{tabular}{|c|c|c|c|}
			\hline 
			\hline 
			 & $ \tilde \zeta _4^\perp $ & $ \kappa _4 ^\parallel $ & $ \kappa _4 ^\perp $ \\
			\hline 
			$ \rho $ & --0.08 & 0 & 0 \\
			$ K^* $ & --0.05 & --0.025 & 0.013 \\
			\hline 
			\hline
		\end{tabular}
	\end{table}
}
The light quark masses are taken to be zero, namely $ m_u = m_d = 0 $, and the mass of the strange quark at $ \mu = 1 \ \birim{GeV} $ is taken to be $ m_s = 0.137 \ \birim{GeV} $.
\section{Details of calculations in the theoretical part}\label{app:B}
In this section, we present the integrals required in the theoretical analysis. In what follows, $ f(u) $ and $ \mathcal F (\alpha _i) $ denote generic 2- and 3-particle DAs, respectively, and we let $ K _i := K _i(m_Q\sqrt{-x^2})/(\sqrt{-x^2})^i $ and $ K_j := K _j(m_{Q'}\sqrt{-x^2})/(\sqrt{-x^2})^j $.
\prg 
\textit{Terms proportional to $ \mathcal O (\langle G \rangle^0) $:} For the sake of simplicity, we suppress the integral measures, $ \int du \int d^4x \ e^{i(p+\baru q)x}$, on the left-hand side. 
\ba
  K_i K_j f(u) &\to& \frac i4 \frac{16\pi^2}{(2m_Q)^i (2m_{Q'})^j} (M^2)^{i+j} f(\frac 12) e^{-m_V^2/2M^2} \int _{(m_Q+m_{Q'})^2} ^{s_0} ds\ e^{-s/M^2} 
  \carp \int d\alpha\ \alpha^{i-1} (1-\alpha)^{j-1} \delta(s - \frac{m_Q^2}{\alpha} - \frac{m_{Q'}^2}{1-\alpha})
\ea
\ba
  x_\mu K_i K_j f(u) &\to& \frac{-i(2p_\mu + q_\mu)}{M^2} \frac i4 \frac{16\pi^2}{(2m_Q)^i (2m_{Q'})^j} (M^2)^{i+j} f(\frac 12) e^{-m_V^2/2M^2} \int _{(m_Q+m_{Q'})^2} ^{s_0} ds\ e^{-s/M^2} 
  \carp \int d\alpha\ \alpha^{i-1} (1-\alpha)^{j-1} \delta(s - \frac{m_Q^2}{\alpha} - \frac{m_{Q'}^2}{1-\alpha})
\ea
\ba
  x_\mu x_\nu K_i K_j f(u) &\to& \frac{-(2p_\mu+q_\mu) (2p_\nu+q_\nu) - 2M^2 g_{\mu\nu}}{M^4} \frac i4 \frac{16\pi^2}{(2m_Q)^i (2m_{Q'})^j} (M^2)^{i+j} f(\frac 12) e^{-m_V^2/2M^2} 
  \carp \int _{(m_Q+m_{Q'})^2} ^{s_0} ds\ e^{-s/M^2} \int d\alpha\ \alpha^{i-1} (1-\alpha)^{j-1} \delta(s - \frac{m_Q^2}{\alpha} - \frac{m_{Q'}^2}{1-\alpha})
\ea
\ba
  x^2 K_i K_j f(u) &\to& \frac i4 \frac{16\pi^2}{(2m _Q)^i (2m_{Q'})^j} M^2 e^{-m_V^2/2M^2} f(\frac 12) \int _{(m_Q + m_{Q'})^2} ^{s_0} ds\ e^{-s/M^2} 
  \carp \int d\alpha \ \alpha^{i-1} (1-\alpha)^{j-1} \delta(s-\frac{m_Q^2}{\alpha} - \frac{m_{Q'}^2}{1-\alpha}) (M^2)^{-3+i+j} 
  \carp \frac{-4(-1+\alpha) (M^2 (-1+i+j) \alpha + m_Q^2) + 4\alpha m_{Q'}^2}{\alpha(\alpha-1)}
\ea
\ba
  x_\mu x^2 K_i K_j f(u) &\to& \frac{-i(2p_\mu+q_\mu)}{M^2} \frac i4 \frac{16\pi^2}{(2m _Q)^i (2m_{Q'})^j} M^2 e^{-m_V^2/2M^2} f(\frac 12) \int _{(m_Q + m_{Q'})^2} ^{s_0} ds\ e^{-s/M^2} 
  \carp \int d\alpha \ \alpha^{i-1} (1-\alpha)^{j-1} \delta(s-\frac{m_Q^2}{\alpha} - \frac{m_{Q'}^2}{1-\alpha}) (M^2)^{-3+i+j} 
  \carp \frac{-4(-1+\alpha) (M^2 (-1+i+j) \alpha + m_Q^2) + 4\alpha m_{Q'}^2}{\alpha(\alpha-1)}
\ea
\ba
  x_\mu x_\nu x^2 K_i K_j f(u) &\to& \frac{-(2p_\mu+q_\mu) (2p_\nu + q _\nu)-2M^2 g_{\mu\nu}}{M^4} \frac i4 \frac{16\pi^2}{(2m _Q)^i (2m_{Q'})^j} M^2 e^{-m_V^2/2M^2} f(\frac 12) 
  \carp \int _{(m_Q + m_{Q'})^2} ^{s_0} ds\ e^{-s/M^2} \int d\alpha \ \alpha^{i-1} (1-\alpha)^{j-1} \delta(s-\frac{m_Q^2}{\alpha} - \frac{m_{Q'}^2}{1-\alpha}) 
  \carp (M^2)^{-3+i+j} \frac{-4(-1+\alpha) (M^2 (-1+i+j) \alpha + m_Q^2) + 4\alpha m_{Q'}^2}{\alpha(\alpha-1)}
\ea
\ba
  x^4 K_i K_j f(u) &\to& \frac i4 \frac{16\pi^2}{(2m_Q)^i (2m_{Q'})^j} M^2 e^{-m_V^2/2M^2} f(\frac 12) \int _{(m_Q+m_{Q'})^2}^{s_0} ds\ e^{-s/M^2} 
  \carp \int d\alpha\ \alpha^{i-1} (1-\alpha)^{j-1} \delta(s - \frac{m_Q^2}{\alpha} - \frac{m_{Q'}^2}{1-\alpha}) \frac{16(M^2)^{-5+i+j}}{\alpha^2(-1+\alpha)^2} \carp(
  M^4 (-2+i+j) (-1+i+j) (-1+\alpha)^2 \alpha^2 
  \ar (-1+\alpha)^2 m_Q^2 (2M^2(-2+i+j) \alpha + m_Q^2)
  \ek 2(-1+\alpha)\alpha (M^2(-2+i+j)\alpha + m_Q^2) m_{Q'}^2 + \alpha^2 m_{Q'}^4
)
\ea
\ba
  x_\mu x^4 K_i K_j f(u) &\to& \frac{-i(2p_\mu+q_\mu)}{M^2} \frac i4 \frac{16\pi^2}{(2m_Q)^i (2m_{Q'})^j} M^2 e^{-m_V^2/2M^2} f(\frac 12) \int _{(m_Q+m_{Q'})^2}^{s_0} ds\ e^{-s/M^2} 
  \carp \int d\alpha\ \alpha^{i-1} (1-\alpha)^{j-1} \delta(s - \frac{m_Q^2}{\alpha} - \frac{m_{Q'}^2}{1-\alpha}) \frac{16(M^2)^{-5+i+j}}{\alpha^2(-1+\alpha)^2} \carp(
  M^4 (-2+i+j) (-1+i+j) (-1+\alpha)^2 \alpha^2 
  \ar (-1+\alpha)^2 m_Q^2 (2M^2(-2+i+j) \alpha + m_Q^2)
  \ek 2(-1+\alpha)\alpha (M^2(-2+i+j)\alpha + m_Q^2) m_{Q'}^2 + \alpha^2 m_{Q'}^4
)
\ea
\ba
  x_\mu x_\nu x^4 K_i K_j f(u) &\to& \frac{-(2p_\mu+q_\mu) (2p_\nu + q _\nu)-2M^2 g_{\mu\nu}}{M^4} \frac i4 \frac{16\pi^2}{(2m_Q)^i (2m_{Q'})^j} M^2 e^{-m_V^2/2M^2} f(\frac 12) 
  \carp \int _{(m_Q+m_{Q'})^2}^{s_0} ds\ e^{-s/M^2} \int d\alpha\ \alpha^{i-1} (1-\alpha)^{j-1} \delta(s - \frac{m_Q^2}{\alpha} - \frac{m_{Q'}^2}{1-\alpha}) 
  \carp \frac{16(M^2)^{-5+i+j}}{\alpha^2(-1+\alpha)^2} (
  M^4 (-2+i+j) (-1+i+j) (-1+\alpha)^2 \alpha^2 
  \ar (-1+\alpha)^2 m_Q^2 (2M^2(-2+i+j) \alpha + m_Q^2)
  \ek 2(-1+\alpha)\alpha (M^2(-2+i+j)\alpha + m_Q^2) m_{Q'}^2 + \alpha^2 m_{Q'}^4
)
\ea
\textit{Terms proportional to $ \mathcal O (\langle G \rangle^1) $:} For the sake of simplicity, we suppress the integral measures, $ \int du \int d^4x \int \mathcal D\alpha_i \ e^{i(p+(\alpha_1 + u\alpha_3) q)x} $, on the left-hand side.
\ba
  K_i K_j f(u) \mathcal F (\alpha_i) &\to& \frac i4 \frac{16\pi^2}{(2m_Q)^i (2m_{Q'})^j} (M^2)^{i+j} e^{-m_V^2/2M^2} \int _{(m_Q + m_{Q'})^2} ^{s_0} ds\ e^{-s/M^2} 
  \carp\int d\alpha\ \alpha^{i-1} (1-\alpha)^{j-1} \delta(s - \frac{m_Q^2}{\alpha} - \frac{m_{Q'}^2}{1-\alpha}) 
  \carp \int _0^{1/2} d\alpha_1 \int _{1/2-\alpha_1}^{1-\alpha_1} d\alpha_3 \ f(\frac{1/2-\alpha_1}{\alpha_3})\frac{\mathcal F(\alpha _i)}{\alpha_3}
\ea
\ba
  x_\mu K_i K_j f(u) \mathcal F (\alpha_i) &\to& \frac{-i(2p_\mu+q_\mu)}{M^2} \frac i4 \frac{16\pi^2}{(2m_Q)^i (2m_{Q'})^j} (M^2)^{i+j} e^{-m_V^2/2M^2} 
  \carp \int _{(m_Q + m_{Q'})^2} ^{s_0} ds\ e^{-s/M^2} \int d\alpha\ \alpha^{i-1} (1-\alpha)^{j-1} \delta(s - \frac{m_Q^2}{\alpha} - \frac{m_{Q'}^2}{1-\alpha}) 
  \carp \int _0^{1/2} d\alpha_1 \int _{1/2-\alpha_1}^{1-\alpha_1} d\alpha_3 \ f(\frac{1/2-\alpha_1}{\alpha_3})\frac{\mathcal F(\alpha _i)}{\alpha_3}
\ea
\ba
  x_\mu x_\nu K_i K_j f(u) \mathcal F (\alpha_i) &\to& \frac{-(2p_\mu+q_\mu) (2p_\nu + q _\nu)-2M^2 g_{\mu\nu}}{M^4} \frac i4 \frac{16\pi^2}{(2m_Q)^i (2m_{Q'})^j} (M^2)^{i+j} e^{-m_V^2/2M^2} 
  \carp \int _{(m_Q + m_{Q'})^2} ^{s_0} ds\ e^{-s/M^2} \int d\alpha\ \alpha^{i-1} (1-\alpha)^{j-1} \delta(s - \frac{m_Q^2}{\alpha} - \frac{m_{Q'}^2}{1-\alpha}) 
  \carp \int _0^{1/2} d\alpha_1 \int _{1/2-\alpha_1}^{1-\alpha_1} d\alpha_3 \ f(\frac{1/2-\alpha_1}{\alpha_3})\frac{\mathcal F(\alpha _i)}{\alpha_3}
\ea
\ba
  x^2 K_i K_j f(u) \mathcal F (\alpha _i) &\to& \frac i4 \frac{16\pi^2}{(2m_Q)^i (2m_{Q'})^j} e^{-m_V^2/2M^2} M^2 \int _{(m_Q + m_{Q'})^2} ^{s_0} ds \ e^{-s/M^2} 
  \carp \int d\alpha \ \alpha^{i-1} (1-\alpha)^{j-1} \delta (s - \frac{m_Q^2}{\alpha} - \frac{m_{Q'}^2}{1-\alpha}) (M^2)^{-3+i+j} 
  \carp \frac{-4(-1+\alpha)(M^2 (-1+i+j)\alpha + m_Q^2) + 4\alpha m_{Q'}^2}{(-1+\alpha)\alpha} 
  \carp \int _0^{1/2} d\alpha _1 \int _{1/2-\alpha_1}^{1-\alpha_1} d\alpha_3 \ f(\frac{1/2-\alpha_1}{\alpha_3}) \frac{\mathcal F(\alpha _i)}{\alpha_3}
\ea
\ba
  x_\mu x^2 K_i K_j f(u) \mathcal F (\alpha _i) &\to& \frac{-i(2p_\mu+q_\mu)}{M^2} \frac i4 \frac{16\pi^2}{(2m_Q)^i (2m_{Q'})^j} e^{-m_V^2/2M^2} M^2 \int _{(m_Q + m_{Q'})^2} ^{s_0} ds \ e^{-s/M^2} 
  \carp \int d\alpha \ \alpha^{i-1} (1-\alpha)^{j-1} \delta (s - \frac{m_Q^2}{\alpha} - \frac{m_{Q'}^2}{1-\alpha}) (M^2)^{-3+i+j} 
  \carp\frac{-4(-1+\alpha)(M^2 (-1+i+j)\alpha + m_Q^2) + 4\alpha m_{Q'}^2}{(-1+\alpha)\alpha} 
  \carp \int _0^{1/2} d\alpha _1 \int _{1/2-\alpha_1}^{1-\alpha_1} d\alpha_3 \ f(\frac{1/2-\alpha_1}{\alpha_3}) \frac{\mathcal F(\alpha _i)}{\alpha_3}
\ea
\ba
  x_\mu x_\nu x^2 K_i K_j f(u) \mathcal F (\alpha _i) &\to& \frac{-(2p_\mu+q_\mu) (2p_\nu + q _\nu)-2M^2 g_{\mu\nu}}{M^4} \frac i4 \frac{16\pi^2}{(2m_Q)^i (2m_{Q'})^j} e^{-m_V^2/2M^2} M^2 
  \carp \int _{(m_Q + m_{Q'})^2} ^{s_0} ds \ e^{-s/M^2} \int d\alpha \ \alpha^{i-1} (1-\alpha)^{j-1} \delta (s - \frac{m_Q^2}{\alpha} - \frac{m_{Q'}^2}{1-\alpha}) 
  \carp (M^2)^{-3+i+j} \frac{-4(-1+\alpha)(M^2 (-1+i+j)\alpha + m_Q^2) + 4\alpha m_{Q'}^2}{(-1+\alpha)\alpha} 
  \carp \int _0^{1/2} d\alpha _1 \int _{1/2-\alpha_1}^{1-\alpha_1} d\alpha_3 \ f(\frac{1/2-\alpha_1}{\alpha_3}) \frac{\mathcal F(\alpha _i)}{\alpha_3}
\ea
\ba
  x^4 K_i K_j f(u) \mathcal F(\alpha_i) &\to& \frac i4 \frac{16\pi^2}{(2m_Q)^i (2m_{Q'})^j} M^2 e^{-m_V^2/2M^2} \int _{(m_Q+m_{Q'})^2}^{s_0} ds\ e^{-s/M^2} 
  \carp \int d\alpha\ \alpha^{i-1} (1-\alpha)^{j-1} \delta(s-\frac{m_Q^2}{\alpha} - \frac{m_{Q'}^2}{1-\alpha}) \frac{16(M^2)^{-5+i+j}}{(-1+\alpha)^2\alpha^2} \carp(
  M^4(-2+i+j) (-1+i+j) (-1+\alpha)^2 \alpha^2 
  \ar (-1+\alpha)^2 m_Q^2 (2M^2 (-2+i+j) \alpha + m_Q^2) 
  \ek 2(-1+\alpha) \alpha (M^2 (-2+i+j) \alpha +m_Q^2) m_{Q'}^2 + \alpha m_{Q'}^4 
) 
	\carp\int _0^{1/2} d\alpha _1 \int _{1/2-\alpha_1}^{1-\alpha_1} d\alpha_3\ f(\frac{1/2-\alpha_1}{\alpha_3}) \frac{\mathcal F(\alpha_i)}{\alpha_3}
\ea
\ba
  x_\mu x^4 K_i K_j f(u) \mathcal F(\alpha_i) &\to& \frac{-i(2p_\mu+q_\mu)}{M^2} \frac i4 \frac{16\pi^2}{(2m_Q)^i (2m_{Q'})^j} M^2 e^{-m_V^2/2M^2} \int _{(m_Q+m_{Q'})^2}^{s_0} ds\ e^{-s/M^2} 
  \carp \int d\alpha\ \alpha^{i-1} (1-\alpha)^{j-1} \delta(s-\frac{m_Q^2}{\alpha} - \frac{m_{Q'}^2}{1-\alpha}) \frac{16(M^2)^{-5+i+j}}{(-1+\alpha)^2\alpha^2} \carp (
  M^4(-2+i+j) (-1+i+j) (-1+\alpha)^2 \alpha^2 
  \ar (-1+\alpha)^2 m_Q^2 (2M^2 (-2+i+j) \alpha + m_Q^2) 
  \ek 2(-1+\alpha) \alpha (M^2 (-2+i+j) \alpha +m_Q^2) m_{Q'}^2 + \alpha m_{Q'}^4 
) 
	\carp \int _0^{1/2} d\alpha _1 \int _{1/2-\alpha_1}^{1-\alpha_1} d\alpha_3\ f(\frac{1/2-\alpha_1}{\alpha_3}) \frac{\mathcal F(\alpha_i)}{\alpha_3}
\ea
\ba
  x_\mu x_\nu x^4 K_i K_j f(u) \mathcal F(\alpha_i) &\to& \frac{-(2p_\mu+q_\mu) (2p_\nu + q _\nu)-2M^2 g_{\mu\nu}}{M^4} \frac i4 \frac{16\pi^2}{(2m_Q)^i (2m_{Q'})^j} M^2 e^{-m_V^2/2M^2} 
  \carp \int _{(m_Q+m_{Q'})^2}^{s_0} ds\ e^{-s/M^2} \int d\alpha\ \alpha^{i-1} (1-\alpha)^{j-1} \delta(s-\frac{m_Q^2}{\alpha} - \frac{m_{Q'}^2}{1-\alpha}) 
  \carp \frac{16(M^2)^{-5+i+j}}{(-1+\alpha)^2\alpha^2} (
  M^4(-2+i+j) (-1+i+j) (-1+\alpha)^2 \alpha^2 
  \ar (-1+\alpha)^2 m_Q^2 (2M^2 (-2+i+j) \alpha + m_Q^2) 
  \ek 2(-1+\alpha) \alpha (M^2 (-2+i+j) \alpha +m_Q^2) m_{Q'}^2 + \alpha m_{Q'}^4 
) 
	\carp \int _0^{1/2} d\alpha _1 \int _{1/2-\alpha_1}^{1-\alpha_1} d\alpha_3\ f(\frac{1/2-\alpha_1}{\alpha_3}) \frac{\mathcal F(\alpha_i)}{\alpha_3}
\ea
For terms containing $ q\cdot x $, we perform the following operation:
\ba 
	(q\cdot x)^n f(u) \to \begin{cases}
		(-i \frac{\del}{\del u})^n f(u), & n > 0\\
		i^{-n} \int _0^u dv _n \cdots \int _0^{v_3} dv_2 \int _0^{v_2}dv_1 f(v_1), & n < 0
	\end{cases}
\ea 
\ba 
	(q\cdot x)^n \mathcal F (\alpha _i) \to \begin{cases}
		(\frac iu \frac{\del}{\del \alpha _3})^n \mathcal F (\alpha _i), & n > 0\\
		(-iu)^{-n} \int _0^{\alpha _3} d\alpha _3^{(n)} \cdots \int _0^{\alpha_3^{(3)}}d\alpha_3^{(2)} \int _0^{\alpha _3 ^{(2)}}d\alpha_3^{(1)} \mathcal F (\alpha _1, 1-\alpha_1-\alpha_3^{(1)},\alpha_3^{(1)}), & n < 0
	\end{cases}
\ea 
\bibliography{paper}
\end{document}